\documentclass{article}
    \usepackage[margin=0.7in]{geometry}
    \usepackage[parfill]{parskip}
    \usepackage[utf8]{inputenc}
    \usepackage{float}
    \usepackage{subcaption}
    \usepackage{caption}
    \usepackage{color}
    \usepackage{soul}
    \usepackage{authblk}

\usepackage{amsmath,amssymb,amsfonts,amsthm,bm,graphicx,epstopdf-base}
    \usepackage{cite}
    \graphicspath{{./Figures/}}

\begin{document}
\title{Exploring the link between coffee matrix microstructure and flow properties using combined X-ray microtomography and smoothed particle hydrodynamics simulations}
\author[1,2]{Chaojie Mo\footnote{cmo@bcamath.org}}
\author[3]{Richard Johnston}
\author[4]{Luciano Navarini}
\author[4]{Furio Suggi Liverani}
\author[1,5,6]{Marco Ellero}
\affil[1]{Basque Center for Applied Mathematics (BCAM), Alameda de Mazarredo 14, 48009 Bilbao, Spain}
\affil[2]{Aircraft and Propulsion Laboratory, Ningbo Institute of Technology, Beihang University, Ningbo 315100, P. R. China}
\affil[3]{Faculty of Science and Engineering, Swansea University, Swansea, SA1 8EN, UK}
\affil[4]{Illycaffè S.p.A, Via Flavia 110, Trieste 34147, Italy}
\affil[5]{Zienkiewicz Centre for Computational Engineering (ZCCE), Swansea University, Bay Campus, Swansea SA1 8EN, UK}
\affil[6]{IKERBASQUE, Basque Foundation for Science, Calle de María Díaz de Haro 3, 48013 Bilbao, Spain}
\date{}
\maketitle
\section*{Abstract}
Coffee extraction involves many complex physical and transport processes extremely difficult to model. Among the many factors that will affect the final quality of coffee, the microstructure of the coffee matrix is one of the most critical ones. In this article, we use X-ray micro-computed (microCT) technique to capture the microscopic details of coffee matrices at particle-level and perform fluid dynamics simulation based on the smoothed particle hydrodynamics method (SPH) with the 3D reconstructured data. Information like flow permeability and tortuosity of the matrices can be therefore obtained from our simulation. We found that inertial effects can be quite significant at the normal pressure gradient conditions typical for espresso brewing, and can provide an explanation for the inconsistency of permeability measurements seen in the literature. Several types of coffee powder are further examined, revealing their distinct microscopic details and resulting flow features. By comparing the microCT images of pre- and post-extraction coffee matrices, it is found that a decreasing porosity profile (from the bottom-outlet to the top-inlet) always develops after extraction. This counterintuitive phenomenon can be explained using a pressure-dependent erosion model proposed in our prior work. Our results reveal not only some important hydrodynamic mechanisms of coffee extraction, but also show that microCT scan can provide useful microscopic details for coffee extraction modelling. MicroCT scan establishes the basis for a data-driven numerical framework to explore the link between coffee powder microstructure and extraction dynamics, which is the prerequisite to study the time evolution of both volatile and non-volatile organic compounds and then the flavour profile of coffee brews.

\section{Introduction}
A roasted and ground coffee matrix is a highly complex porous medium characterized by  multiscale features. A typical coffee matrix used for double-shot espresso brew is about 15 ml in volume and consists of millions of coffee particles. These coffee particles are produced by grinding roasted coffee beans, and have an approximately bimodal size distribution\cite{Povey_IOP_2020}. The first peak of the distribution profile is always at around $30\sim 40\ \mathrm{\mu m}$, representing the so-called ``fines'', i.e.,\ inner walls cellular fragments (to be very rigorous, this first peak is preceded by a small submicron fraction). The second peak represents the so-called ``coarses'' and their mean size depends on the grinder, with values ranging between $200 \sim 1000\ \mathrm{\mu m}$. Moreover, the coffee particles are not entirely solid but porous material themselves, characterized by many cell-pockets of dimensions $30 \sim 60\ \mathrm{\mu m}$ inside the particles\cite{Povey_IOP_2020}. Finally, the cell-walls are also porous at nanoscale level.

During coffee extraction, as water flows through this complex porous structure,
many chemical compounds including among many others caffeine, trigonelline, polyphenols and
polysaccharides, are diffused and convected out of the solid matrix to finally enter into the cup and determine the quality (e.g. flavour and mouthfeel) of the final beverage. 

This coffee extraction process is especially difficult to model. This is not simply because of the complex topological features related to the multiscale nature of coffee bed geometry as explained above, there are also many other physical processes concomitant to the simple diffusion and convection. For example, the porous medium undergoes geometric changes due to the fine migration\cite{Petracco1993}, the particle erosion\cite{Hargarten2020,Matias2020}, and particle swelling\cite{Hargarten2020,Maille2021}; there is also solubilization of many hydrophilic substances\cite{Voilley1980}, the emulsification of insoluble coffee oils\cite{Illy2005}, suspension of solid coffee cell-wall fragments (fines)\cite{Illy2005}, and $\mathrm{CO}_2$ degassing and supersaturation\cite{Illy2011}. These physical processes are all interconnected, rendering the modelling of coffee extraction very challenging.

In order to predict the coffee quality and propose better brewing protocol \textit{a priori}, many works have tried to model the extraction process using a variety of simulation techniques. We mention early attempts using cellular automata based model for percolation simulation\cite{Petracco1993,Cappuccio2001}. More recently, Lattice Bolztmann Method (LBM) has been also used to simulate the extraction kinetics\cite{Rosa2016} and to consider the swelling and erosion effects\cite{Matias2020}. Multiscale models have also been proposed. In the work of Moroney \textit{et al}\cite{Moroney2015} an upscaling procedure is applied to relate the conservation equations at different scales. The developed multiscale model is able to quantitatively reproduce the experimental extraction profile after being parameterized by experimentally measured coffee bed parameters. In the work of Melrose \textit{et al}\cite{Melrose2018} and Cameron \textit{et al}\cite{Cameron2020}, another multiscale model, which can incorporate the bimodal characteristics of the coffee particle, is proposed. In this model the macroscopic mass transport is modelled using a 1-dimensional convection-diffusion equation, while the microscopic extraction is modelled through diffusion equation inside representative coffee particles in different layers of the coffee matrix. The equations at different scale are coupled through source terms and boundary conditions. This model is also found to be able to capture the extraction kinetics. Besides these models, mesoscopic model based on smoothed dissipative particle dynamics (SDPD) has been also recently proposed to incorporate the fine migration\cite{Ellero2019}, particle erosion\cite{Mo2021} and swelling effect\cite{Mo2022} during coffee extraction.

However, in most of the aforementioned modelling and simulations, the fluid dynamics through the coffee matrix is either characterized by average porosity and permeability\cite{Rosa2016,Moroney2015,Melrose2018,Cameron2020} or is resolved by flows through artificial packing of many spheres/disks\cite{Ellero2019,Matias2020,Mo2021,Mo2022}. These simulations may not be able to capture some important features of coffee extraction. For example, the coffee particles size is usually of smooth bimodal distribution\cite{Povey_IOP_2020} and the particles are not spheric\cite{Hargarten2020}, therefore many details are lost if the simple Carman-Kozeny relation and Darcy's law are used to characterize the flow, or if some synthetic porous medium formed by packing ideal spheres/disks are used for percolation simulations. In fact, it is known that the particle size distribution is quite influential to the final quality of coffee\cite{Kuhn2017}. Therefore, it is important to accurately model the whole spectrum of particle size in coffee extraction simulation. A straightforward high-resolution mapping - capturing all the microscopic details of a coffee matrix - can be obtained by using X-ray microtomography (microCT)/X-ray microscopy (XRM) to scan the coffee matrix, and build a porous medium model using the 3D-reconstructured inner topology\cite{DOnofrio2017,Burridge2019}. In this way, not only the full particles size distribution can be considered, but the spatial permeability fluctuation in the matrix can also be captured. The spatial permeability fluctuation could correlate to the ``partially clogged'' phenomenon which may cause very fine coffee powder to have less extraction rate\cite{Cameron2020}. 

With X-ray microCT, it is also possible to scan both pre-extraction (dry) coffee matrix and post-extraction (wet) coffee matrix. The comparison could reveal some information about how the porous structure changes during extraction process. This information could provide important guidance to model different physical processes that lead to geometric change and matrix restructuring during extraction. As far as we are able to determine, microCT reconstruction of coffee matrix microstructure has not been used for direct coffee extraction simulation. This will be the subject of the current article.

In this article, we will use microCT reconstructured data to build a static porous medium model, and simulate the percolation process using the smoothed particle hydrodynamics method (SPH). We determine the simulated medium permeability at different pressure gradient and compare them to experimental data in literatures. We show that inertial effects are quite influential at the pressure gradients typically considered in espresso brewing. As a result, the inconsistency of many results of coffee matrix permeability found in literature can be explained by the above-mentioned inertial effect. Several types of coffee powder are also examined, and their permeability and tortuosity are measured systematically using SPH simulation. This allows to establish a link between coffee powder granulometry and other microstructural features with extraction efficiency. Finally, microCT images of post-extraction coffee matrices are also analyzed, it is revealed that a decreasing porosity profile (from the bottom-oulet to the top-inlet) always develops after extraction. This counterintuitive phenomenon can be explained using a specific pressure-dependent erosion model proposed in our previous work\cite{Mo2021}.

The article is organized as follows: In section \ref{sec:materials_and_methods} we introduce the coffee powders we study (section \ref{sec:coffee_powders}), the particle size analysis method (section \ref{sec:particle_size_analysis}), the microtomography techinique (section \ref{sec:microtomography}), and the SPH-based percolation simulation technique (section \ref{sec:simulation_using_SPH}). In section \ref{sec:results_and_discussion}, the powder granulometry is first presented (section \ref{sec:granulometry}). Then we show a brief analysis of the microCT images in section \ref{sec:images_analysis}. Discussion of percolation and tortuosity of the coffee matrix is done in section \ref{sec:permeability}. The analysis of the post-extraction microCT images is reported in section \ref{sec:post-extraction}. Finally conclusions are drawn in section \ref{sec:conclusions}.

\section{Materials and methods}
\label{sec:materials_and_methods}
\subsection{Coffee powders}
\label{sec:coffee_powders}
Four different types of ground coffee obtained by the same medium roasted 100\% Coffea
arabica blend were used: Espresso (type E) characterized by the typical size distribution for traditional
espresso brew preparation, IPSO (type H) typical for capsule espresso brew preparation , Refilly Moka (type M) typical for moka-pot preparation, and PillowPack (type F) typical for drip-filter preparation. In the case of type H dark roasting degree was also used. The coffee powders have been packaged in
single dose capsule for espresso brew (Iperespresso\textsuperscript{\textregistered}). The capsule (single shot espresso) is approximately cylindrical
around 14.6 mm in height by 32.5 mm in width. Each capsule contains 6.7$\pm$0.1 gram of
coffee powder. The capsules used in our experiment are sent by illycaffè S.p.A.

\subsection{Particle size analysis}
\label{sec:particle_size_analysis}
The powders are examined with laser diffraction to obtain the particle size characteristics. A Mastersizer 3000 (Malvern Instrument, UK) is used.

\subsection{X-ray microtomography}
\label{sec:microtomography}

X-ray 3D imaging is used to obtain the internal microscopic structure of the coffee matrix. A Zeiss Xradia Versa 520 (Carl Zeiss XRM, Pleasanton, CA, USA) was used to carry out high-resolution XRM non-destructive imaging; this was achieved using a CCD (charge coupled device) detector system with scintillator-coupled visible light optics and a tungsten transmission target. 3D scans were performed with an X-ray tube voltage of 80 kV, a tube current of 87 mA, an exposure of 3000 ms. A total of 3201 projections were collected. An objective lens giving an optical magnification of 0.4x was selected with binning set to 1, producing an isotropic voxel (three-dimensional pixel) size of around $16.8 \mathrm{\mu m} \times 16.8 \mathrm{\mu m} \times 16.8 \mathrm{\mu m}$. The tomograms were reconstructed from 2D projections using a Zeiss Microscopy commercial software package (XMReconstructor) and an automatically generated cone-beam reconstruction algorithm based on the filtered back-projection. The powder are in Illy Hyper Espresso capsule (see detailes above).

\begin{figure}[hbt]
\centering
  \includegraphics[width=0.8\linewidth]{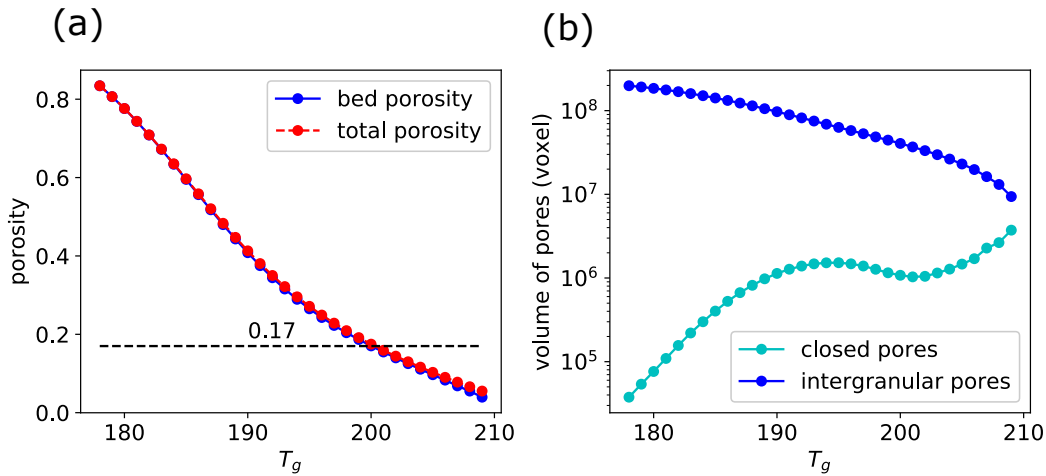}
  \caption{(a) Bed porosity and total porosity of the cuboid sample at different grayscale thresholds. The dash black line mark the typical bed porosity for a realistic coffee matrix. (b) Volume of the intergranular pores and the intragranular closed pores at different grayscale thresholds.}
  \label{fig:Tg_porosity}
\end{figure}

Segmentation is acquired by choosing a global greyscale threshold to binarise the microCT imaging. However, the resolution of the microCT scan is of the same order as the smaller fines, and some of these fines are occupying the air space, therefore the greyscale of the air space is actually a function of the partial volume of air and fines. Furthermore, the coffee particles also have many internal pores of the same scale as the microCT resolution. So the grayscale of the coffee is also a function of the partial volume of solid and air inside the coffee particles. It is therefore quite challenging to determine the proper greyscale threshold distinguishing the coffee and air space. Moreover, the greyscale distribution of the microCT imaging of the coffee bed is unimodal and of Gaussian shape, and conventional segmentation techniques\cite{Baradez2004,Coudray2010} typically fails on these images. Here we determine the greyscale threshold by matching the bed porosity from the microCT imaging and that from  Pycnometer data. The procedure is as follows:
\begin{enumerate}
	\item The microCT images are inverted. In the original image brighter pixels represent solid. We first invert the values of the pixels.
	\item Assuming the greyscale threshold is known as $T_g$, the microCT imaging are binarized using this threshold.
	\item Then a connected-region labeling algorithm\cite{itseez2015opencv} is used to identify the connected and isolated voids.  These clusters of void will consist of one major cluster with very large volume and many minor clusters with small volume. Apparently, the major cluster represents the inter-granular pores. The isolated minor clusters are closed pores inside the grains. We can get the bed porosity, closed porosity, and the total porosity from the volume of these clusters.
    \item To avoid dealing with the cylindrical surface of the coffee cake, we extract a cuboid of $600\times600\times600 \mathrm{voxels}$ ($1\times 1\times 1 \mathrm{cm}$) from each of the microCT imaging of the coffee cakes. Varying $T_g$, the porosity of this cuboid sample at different threshold can be determined. An example obtained by processing the microCT imaging of type E powder is shown in Fig.~\ref{fig:Tg_porosity}. As can be seen from the figure, the total porosity is almost the same as the bed porosity, the total volume of the intergranular pores is also orders higher than the intragranular closed pores, indicating that most of the closed pores are not captured in our microCT imaging. This is reasonable since the resolution of the microCT imaging is $16.759\mathrm{\mu m^3/voxel}$ and the intragranular pores are of the same scale. The resolution is not high enough to accurately capture the intragranular pores. Therefore the porosity consists overwhelmingly of the bed porosity of the coffee cake. By matching the bed porosity obtained from binarized images with the typical bed porosity of coffee matrix available in literatures\cite{Roman-Corrochano2015,Cameron2020}, we can determine the correct grayscale threshold and use that threshold to binarize the microCT imaging. In the example presented in Fig.~\ref{fig:Tg_porosity}, the porosity to be matched is 0.17, the grayscale threshold can be determined to be 200. Then this grayscale threshold can be used to binarize the microCT image.
\end{enumerate}

\subsection{Percolation simulation using SPH}
\label{sec:simulation_using_SPH}

After the segmentation, we have a label matrix from the voxel model, with the solid voxels labelled by 1, and void/air voxels labeled by 0. For the simulations we use Smoothed Particle Hydrodynamics (SPH) method (see Supplementary Material). A simulation box of specified size is filled with SPH particles in specified number density and in some lattice form. By mapping this simulation box to a 3-D region inside the voxel model, the label matrix can be interpolated to classify each SPH particle as fluid particle or solid particle. The construction of SPH particle model from microCT imaging is finished after the classification.

\begin{figure}[!htb]
     \centering
     \includegraphics[width=.5\linewidth]{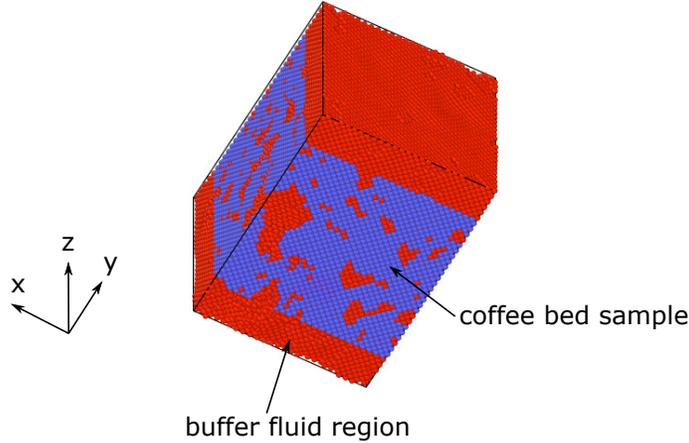}
     \caption{An example of SPH particle model with $L_x = L_z = 708\mathrm{\mu m}$, $L_y=991\mathrm{\mu m}$. Blue particles: solid grains, red particles: fluid.}
     \label{fig:SPH_model}
\end{figure}

An example of SPH particle model generated with the above procedure is shown in Fig.~\ref{fig:SPH_model}. In this example a cube of length $L = 708\mathrm{\mu m}$ is extracted from the microCT imaging and used for the above procedure. Buffer fluid regions at both the top and bottom have been added. The SPH particle density is specified to be the same as the number density of the voxels, and the particles are at face-centered cubic lattice points. 

With the SPH particle model ready, we can set periodic boundary condition at all boundaries, and applied body force on the $y$ direction to drive fluid to percolate through the porous medium. The readers are referred to the Supplementary Material for the details of the SPH formalism and the SPH parameters used in our simulations.

\section{Results and discussion}
\label{sec:results_and_discussion}
\subsection{Coffee powder granulometry}
\label{sec:granulometry}
\begin{figure}[hbt]
\centering
  \includegraphics[width=0.9\linewidth]{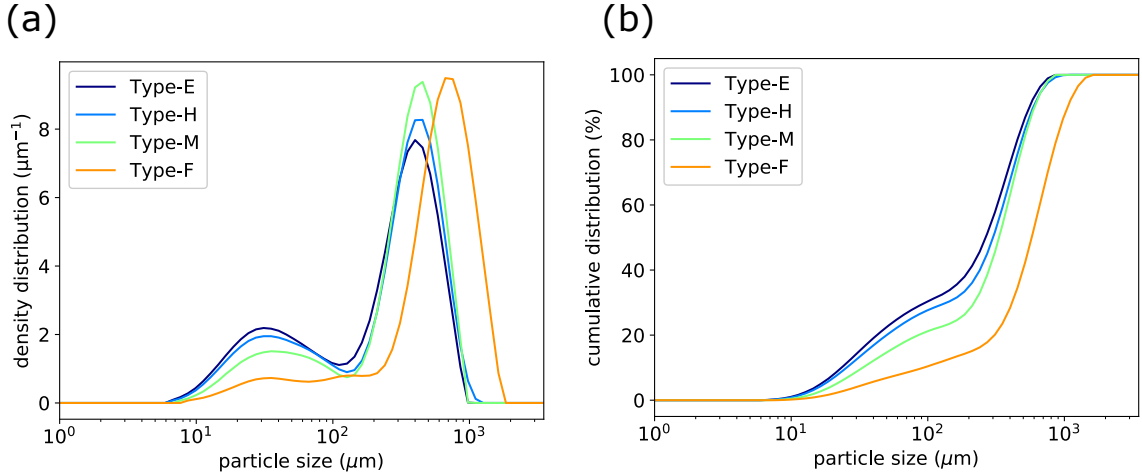}
  \caption{(a) Density distribution of the particle size for different types of ground coffee. (b) Cumulative distribution of the particle size.}
  \label{fig:distribution}
\end{figure}

The particle distributions for the coffee powders measured with laser diffraction are shown in Fig.~\ref{fig:distribution}. The density distributions are typically bimodal, with the first peak representing the fine particles and the second peak the coarse particles. The first peak is always around $25\sim 50\mathrm{\mu m}$, because the fines are cell fragments and the cell size is in the range $25\sim 50\mathrm{\mu m}$. While the position of the coarse peak will shift right with coarser grind.
More results from the laser diffraction measurement, including the uniformity, specific surface area, Sauter diameter (d[3,2]), De Brouckere mean diameter (d[4,3]), and the volume fraction of particles below $100\mathrm{\mu m}$, are summarised in Tab.~\ref{tab:tab0}. It can be seen from both Fig.~\ref{fig:distribution} and Tab.~\ref{tab:tab0} that from type E to type F, the volume fraction of fines is decreasing. If we assume that all particles smaller than $\mathrm{100 \mu m}$ are fines then there are 29.21\% of fines in type E powder but only 9.7\% of fines in type F powder. The mean diameter also increases from type E to type F, and the uniformity and specific surface area decreases concomitantly.

  \begin{table}
	\centering
	\caption{Results from laser diffraction on different types of coffee powder}
	\label{tab:tab0}
	\begin{tabular}{l|ccccc}
    Type & Uniformity & Specific Surface Area ($\mathrm{m^2/kg}$) & d[3,2] $\mathrm{\mu m}$ & d[4,3] $\mathrm{\mu m}$ & Volume below 100 $\mathrm{\mu m}$ \\
    \hline
    E & 0.646 & 74.09 & 81.0 & 308 & 29.21\% \\
    H & 0.600 & 66.60 & 90.1 & 341.6 & 27.52\% \\
    M & 0.508 & 51.66 & 116 & 372 & 20.26\% \\
    F & 0.473 & 28.05 & 214 & 673 & 9.7\% 
	\end{tabular}	
  \end{table}

\subsection{Xray microCT analysis}
\label{sec:images_analysis}
\begin{figure}[hbt]
\centering
  \includegraphics[width=0.5\linewidth]{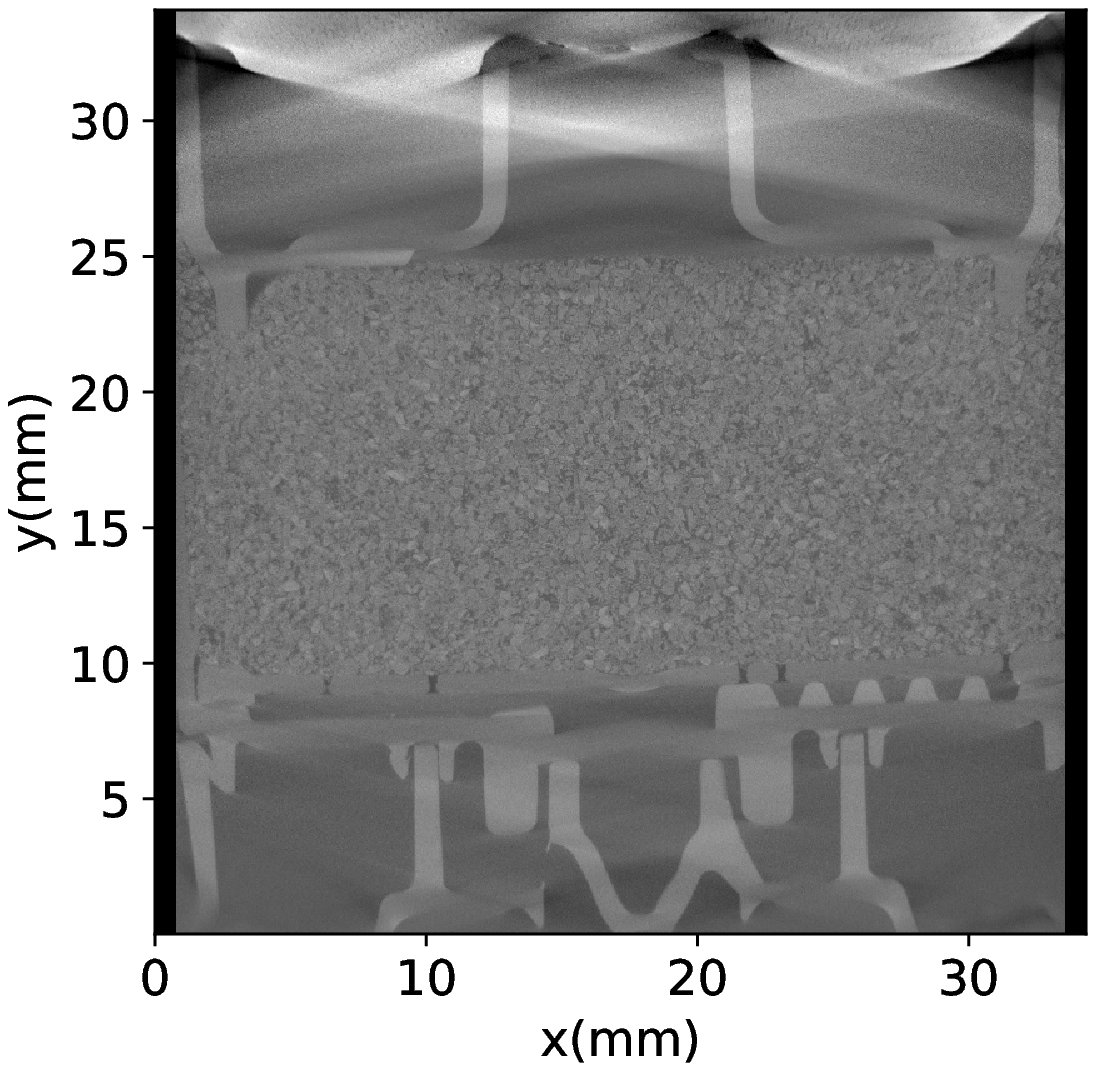}
  \caption{A vertically-cut section of the microCT imaging for the type H coffee powder.}
  \label{fig:typeH_CT}
\end{figure}
A vertical digital section of the microCT/XRM imaging data of type H powder is shown in Fig.~\ref{fig:typeH_CT} (Enlarged cutting section images for all the powders are provided in the Supplementary Materials). Both the coffee grains and the plastic/metal structure skeletons of the capsule are clearly visible in the image. 

\begin{figure}[hbt]
\centering
  \includegraphics[width=0.6\linewidth]{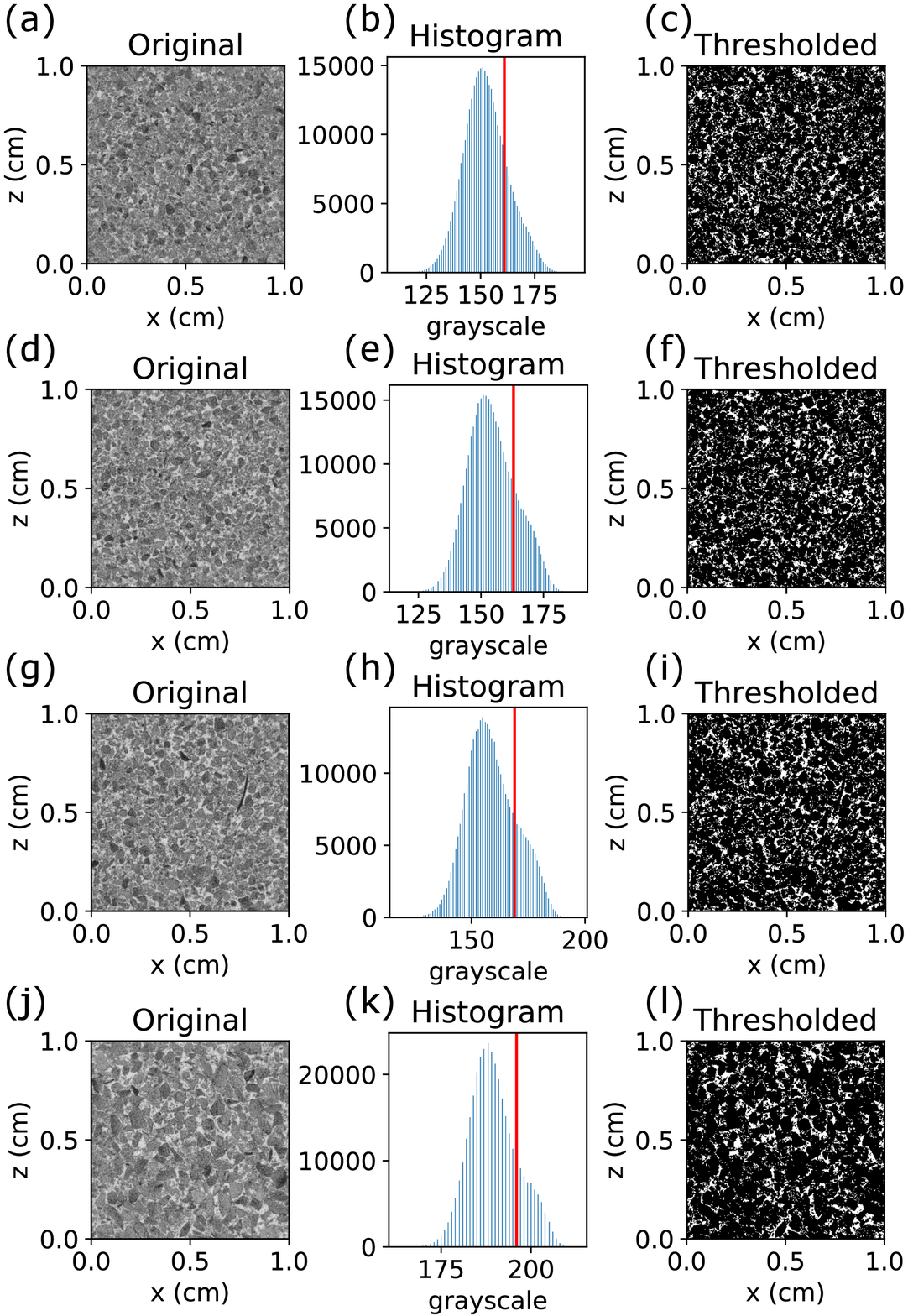}
  \caption{Cropped horizontally-cut sections of the microCT imaging before and after segmentation. (a)-(c) Type E powder; (d)-(f) Type H powder; (g)-(i) Type M powder; (j)-(l) Type F powder. The first column shows the original microCT images. The second column shows the histogram of the grayscale. The red lines mark the grayscale thresholds determined by matching the bed porosity. The third column shows the thresholded images.}
  \label{fig:thresholded}
\end{figure}

The microCT imaging is first binarised following the segmentation protocol introduced above. We show some cropped sections of the microCT imaging before and after segmentation in Figure \ref{fig:thresholded}. Note that in this figure the images before segmentation have already had their pixels inverted so that brighter regions represent voids. As can be seen from the images, the coffee grains are becoming coarser and coarser from type E to type F. This is in agreement with our particle size measurement.

\begin{figure}[hbt]
\centering
  \includegraphics[width=0.9\linewidth]{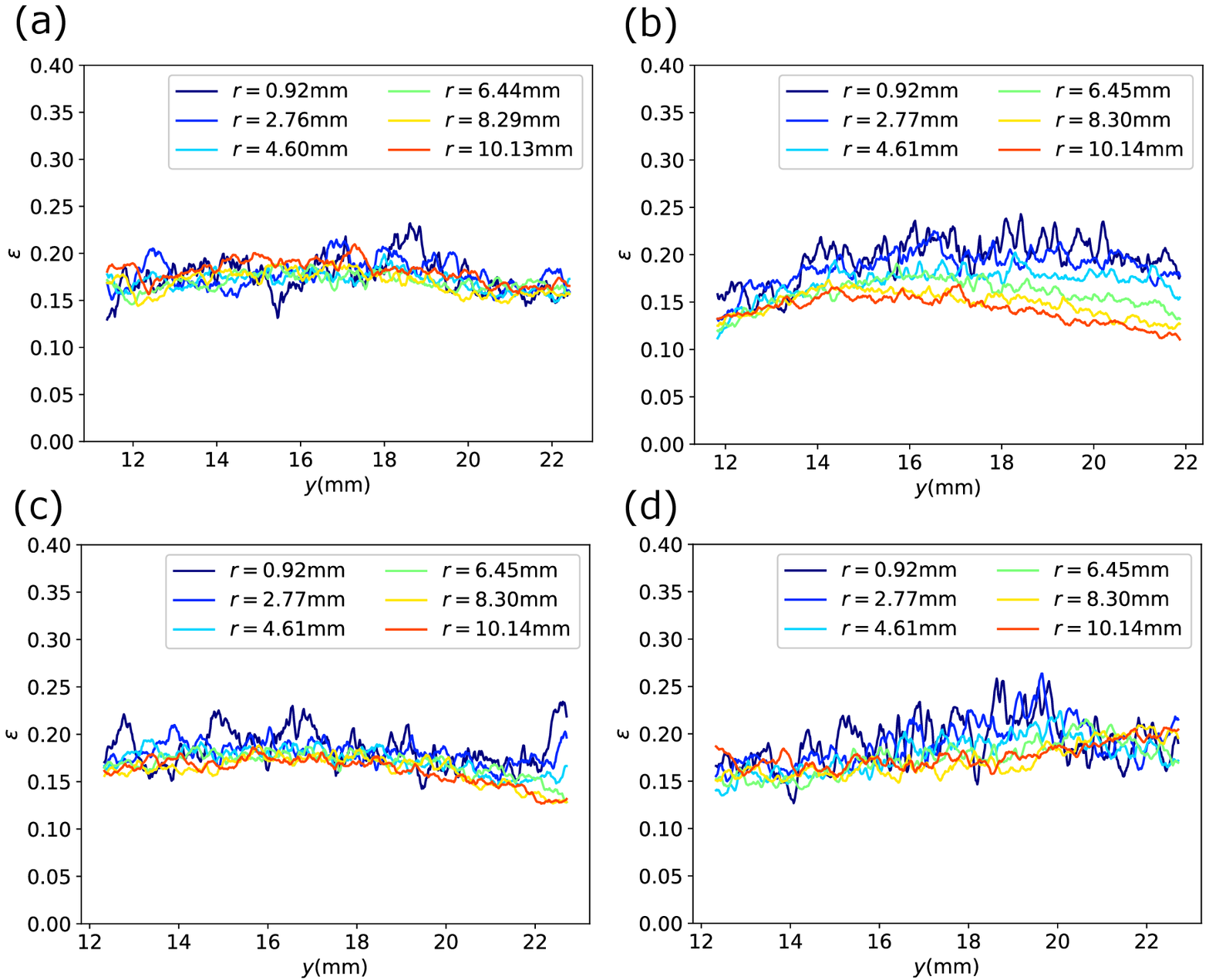}
  \caption{Local porosity of the coffee cakes at different position. $y$ is the vertical position, $r$ is the radial distance to the center of the cake. (a) Type E; (b) Type H; (c) Type M; (d) Type F.}
  \label{fig:porosity_y}
\end{figure}

Then the binarised images are used  to analyse the spatial distribution of porosity of the coffee cakes. Because the coffee cakes are of cylinder shape, we divide the cakes into many annular section with the same radial thickness and calculate the volume fraction of the void in each vertical layer, so that the variation curves of porosity against the vertical coordinate at different radial distance to the centre are acquired as shown in Fig.~\ref{fig:porosity_y}. It can be seen that with the decreasing volume fraction of fines, from (a) to (d) the curves fluctuates more significantly, but in each capsule the porosity is relatively homogenous across the matrix. In most cases, there is no significant variation trend in the radial direction. Except that for type H capsule, there is an apparent trend of higher porosity near the top center. This could be a result of different transportation condition for the capsule. In all cases, there is no significant variation trend in the vertical direction.

\subsection{Permeability and tortuosity}
\label{sec:permeability}
\begin{figure}[!htb]
     \centering
     \includegraphics[width=.5\linewidth]{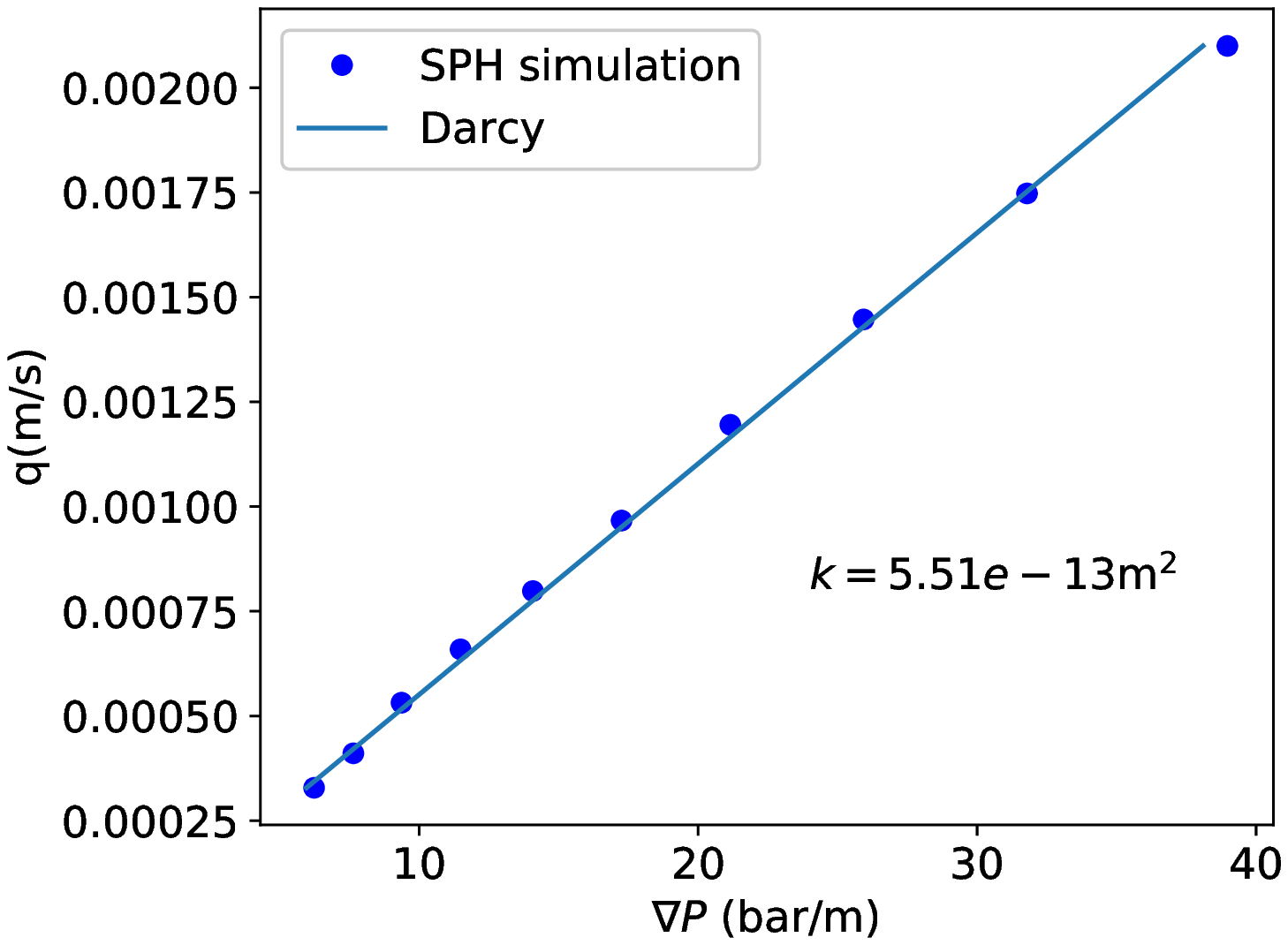}
     \caption{Flow rate at different pressure gradients and the fit using Darcy's law. The pressure gradient is relatively low here. (Type H powder)}
     \label{fig:permeability_LP}
\end{figure}
\begin{figure}[!htb]
     \centering
     \includegraphics[width=.9\linewidth]{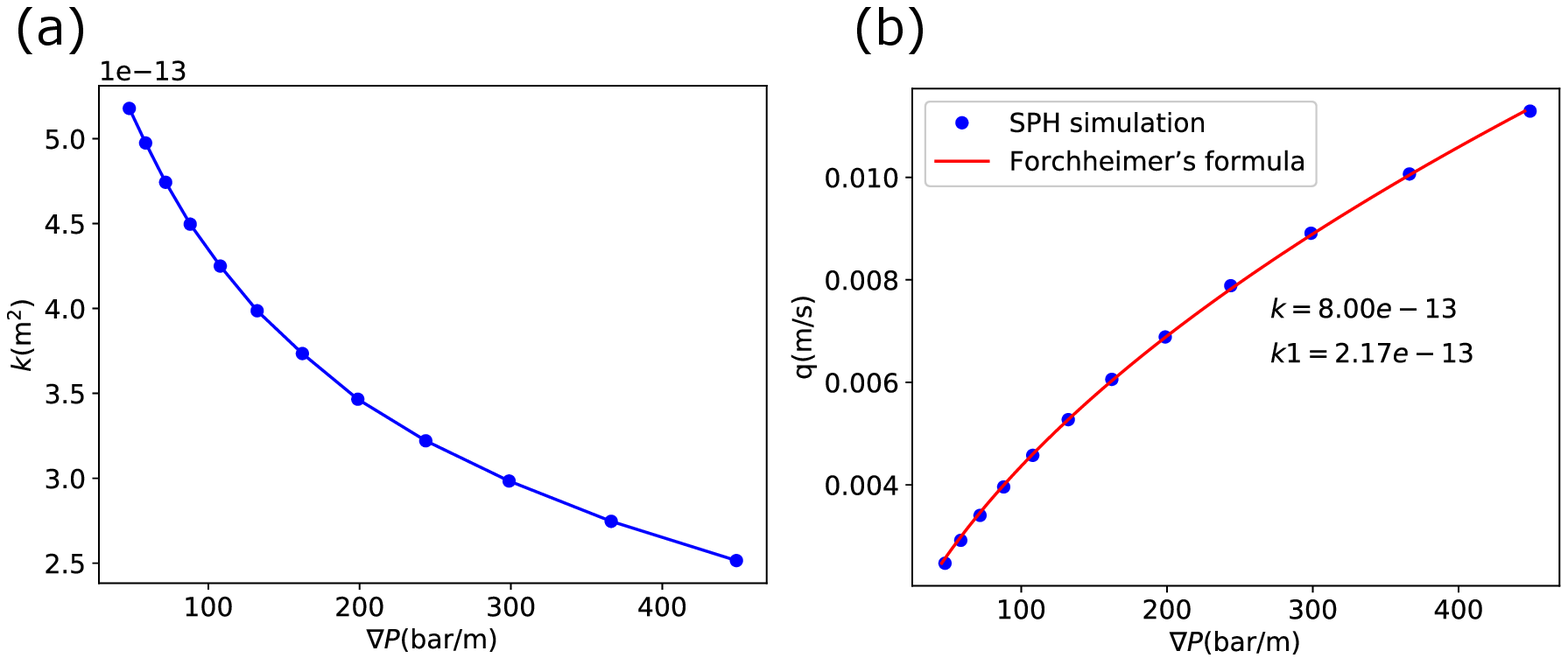}
     \caption{(a) When the pressure gradient is relatively high, the permeability obtained from Darcy's law is dependent on the pressure gradient. (b) Flow rate at different pressure gradients and the fit using Darcy's law with Forchheimer's modification. (Type H powder)}
     \label{fig:permeability_HP}
\end{figure}

As a start, we investigate the permeability of type H powder in detail. The porous flow is simulated with SPH. A cropped cubic sample of length $L = 708\mathrm{\mu m}$ is first examined. The SPH particle model has been shown in Fig.~\ref{fig:SPH_model}. Fluid buffer regions with a total height $L_\mathrm{buf}=283\mathrm{\mu m}$ are added in the $y$ direction. When a body force $g$ is applied in the $-y$ direction, the effective pressure gradient in the sample region will be $\rho g (1+L_\mathrm{buf}/L)$. Figure \ref{fig:permeability_LP} shows the relation of the effective pressure gradient and the resultant flow rate in $-y$ direction. The relation is linear and the fit using Darcy's law:
\begin{equation}
\nabla P = -\frac{\mu}{k}q
\end{equation}
gives a permeability $k=5.51\times 10^{-13}\ \mathrm{m^2}$. This value is smaller than the measurements by Gianino\cite{Gianino2007} ($k = 2.3\times 10^{-12}\ \mathrm{m}^2$) and King\cite{King2008}($k = 8.98\times 10^{-13}\ \mathrm{m}^2$). But it is larger than the estimation given by Navarini \textit{et al}\cite{Navarini2009} ($k=0.7\sim 4\times 10^{-13}\ \mathrm{m^2}$) and Corrochano \textit{et al}\cite{Corrochano_JFE_2015} ($k = 2.59\times 10^{-14}\sim 3.36\times 10^{-13}\ \mathrm{m^2}$). The disagreement is probably caused by the different overpressure ($\Delta P$ across the coffee bed), as in Gianino's\cite{Gianino2007} and King's\cite{King2008} experiments the overpressure is only in the order of $0.01\ \mathrm{bar}$, while in the experiments of Navarini \textit{et al}\cite{Navarini2009} and Corrochano \textit{et al}\cite{Corrochano_JFE_2015} the overpressure is in the order of $1\ \mathrm{bar}$. Here in our SPH model (Fig.~\ref{fig:SPH_model}), the pressure gradient $8\sim 40\  \mathrm{bar/m}$ represents an overpressure between $0.16\ \mathrm{bar}$ and $0.8\ \mathrm{bar}$ on a coffee bed of $2\ \mathrm{cm}$ height. So the overpressure is intermidiate here, resulting in a permeability higher than that of Navarini \textit{et al}\cite{Navarini2009} and Corrochano \textit{et al}\cite{Corrochano_JFE_2015}. When the pressure gradient is significantly increased as shown in Fig.~\ref{fig:permeability_HP} (a) the permeability obtained using Darcy's law decreases with the increasing pressure head, achieving a better agreement with Navarini's\cite{Navarini2009} and Corrochano's\cite{Corrochano_JFE_2015} results. In Fig.~\ref{fig:permeability_HP} (a) the overpressure is in $1\sim 8\ \mathrm{bar}$ when the pressure gradient increase from $50\ \mathrm{bar/m}$ to $400\ \mathrm{bar/m}$ if a coffee bed of $2\ \mathrm{cm}$ height is considered. With the increase of pressure head the permeability decreases from about $5.2\times 10^{-13}$ to $2.5\times 10^{-13}\ \mathrm{m^2}$. This kind of pressure dependence was also reported in Navarini's experimental investigation\cite{Navarini2009}. In the experiment the dependence of permeability on pressure gradient might be attributed to the deformation of the bed under pressure. But in our simulation the porous coffee bed is static and not deformable, and the dependence is still observed, so the dependence can not be all attributed to the deformation of the porous bed under pressure. Here, we believe that the permeability decreases with higher pressure as a result of the flow inertial effects. In Fig.~\ref{fig:permeability_HP} (b), the flow rate at different pressure gradients and the fit using Darcy's law with Forchheimer's modification\cite{Bejan2013} (Eq.~\ref{eq:Forchheimer}) is presented. 
\begin{equation}
\nabla P = -\frac{\mu}{k}q-\frac{\rho}{k_1}|q|q
\label{eq:Forchheimer}
\end{equation}
The fit is very good and gives a permeability ($k=8.0\times 10^{-13} \ \mathrm{m^2}$) and an additional inertial permeability $k_1=2.17\times 10^{-13}\ \mathrm{m^2}$. The inertial effects cause the flow rate to increase in a rate slower than linear relation, so when the Darcy's law is used it gives a permeability decreasing with an increasing pressure gradient. If we use the $d_{4,3}=341.6\ \mathrm{\mu m}$ as the characteristic length, the Reynolds number ($\mathrm{Re}=d_{4,3}\rho q/\mu$) in Fig.~\ref{fig:permeability_HP} is between 0.84 to 3.86, indicating that the inertia is indeed important at high pressure gradient.

Assuming that there is no recirculation, the tortuosity of the excerpt coffee bed sample can be determined by tracing the trajectories of the fluid particles:

\begin{equation}
\tau = \frac{\sum_{i}^N \int_{\mathbf{x}_i\in \Omega_c} ||\mathrm{d}\mathbf{x}_i||}{\left|\sum_{i}^N \int_{\mathbf{x}_i\in \Omega_c} \mathrm{d}\mathbf{x}^\beta_i\right|},
\label{eq:tortuosity}
\end{equation}
where $\tau$ is the tortuosity, $N$ is the total fluid particles, $\Omega_c$ is the coffee sample domain, $\mathbf{x}_i$ is the trajectory of particle $i$, $\beta$ is the direction of the filtration.

\begin{figure}[!htb]
     \centering
     \includegraphics[width=.6\linewidth]{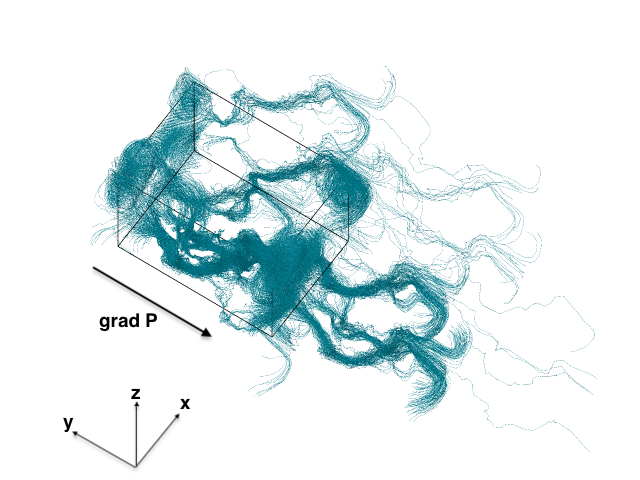}
     \caption{Trajectories of fluid particles through the coffee bed sample. The arrow marks the direction of the pressure gradient. The black frame is the simulation box with dimension $L_x = L_z = 708\mathrm{\mu m}$, $L_y=991\mathrm{\mu m}$, all boundaries are periodic, all coordinates have been unwrapped. (Type H powder)}
     \label{fig:trajectory}
\end{figure}

For one of the simulations in Fig.~\ref{fig:permeability_HP}, the trajectories of 3706 labeled particles whose initial positions are at the top of the upper buffer fluid region, are shown in Fig.~\ref{fig:trajectory}. Because all boundaries are periodic and the coordinates have been unwrapped, the trajectories span outside the simulation box. Using Eq.~\ref{eq:tortuosity} the tortuosity can be determined to be $\tau = 1.80$. The power law relationship has been frequently used to describe the tortuosity-porosity correlation\cite{Dias2006}:
\begin{equation}
\tau = \left( \frac{1}{\epsilon} \right)^{n},
\label{eq:powerlaw}
\end{equation}
where $n$ is a parameter dependent on the packing properties. The porosity of the sample is known to be 0.17, using Eq.~\ref{eq:powerlaw}, it can be determined that $n = -\log\tau / \log\epsilon = 0.33$.

\begin{figure}[!htb]
     \centering
     \includegraphics[width=.7\linewidth]{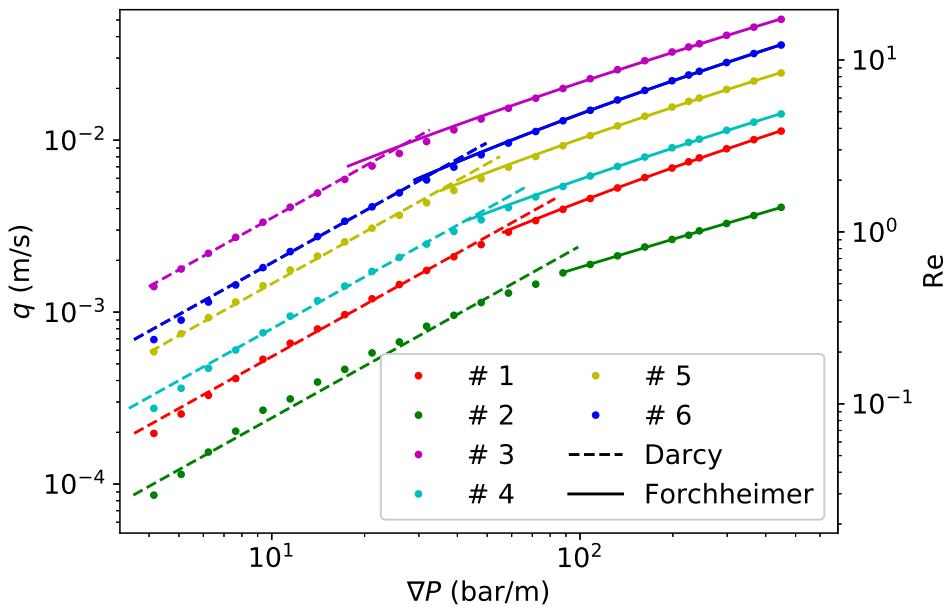}
     \caption{Discharge at different pressure gradient of different excerpt coffee samples. The dash lines are the fittings using Darcy's law, the solid lines are the fittings with Forchheimer's formula. (Type H powder)}
     \label{fig:permeability_all}
\end{figure}

The results shown above are only obtained from a small sample of the whole coffee bed. The bed is however quite inhomogenous, and the porosity can vary significantly across the whole bed in both radial and axial direction as shown in Fig.~\ref{fig:porosity_y}. Therefore, to properly characterize the property of the coffee bed we extract more samples from the bed and perform simulations on them. Results and the locations of the samples are summarized in Tab.~\ref{tab:tab1} and Fig.~\ref{fig:permeability_all}. In Fig.~\ref{fig:permeability_all}, the discharge - pressure gradient relation for different samples have been fitted using both Darcy's law and Forchheimer's modification, in low $\nabla P$ regime and relatively higher $\nabla P$ regime, respectively. All the samples show significant deviation from Darcy's law in the relatively high $\nabla P$ regime. But the Forchheimer's formula accurately captures the $q-\nabla P$ relationship in the high $\nabla P$ regime. Note that in Fig.~\ref{fig:permeability_all} the transitions from Darcy to Forchheimer do not correspond to one single critical Reynolds number. This phenomenon has also been reported by other researchers in the field of hydrology: coarser particles lead to higher critical Re \cite{Li2019}. The permeabilities obtained using Darcy's law in the low $\nabla P$ regime are presented in Tab.~\ref{tab:tab1} as $k^\mathrm{D}$, the permeabilities obtained using Forchheimer's modification in the relatively high $\nabla P$ regime are presented in Tab.~\ref{tab:tab1} as $k^\mathrm{F}$ and $k^\mathrm{F}_1$ for inertia permeability. The porosity and tortuosity for each sample are also presented in Tab.~\ref{tab:tab1}. It can be seen that both the porosity and tortuosity fluctuate slightly, and there is a strong correlation between the porosity and the measured $k^\mathrm{D}$. Higher porosity tends to lead to larger $k^\mathrm{D}$. 
\begin{table}
\centering
\caption{Summary of the properties of the coffee samples (type H powder).  The parameter $k^\mathrm{D}$ is the permeability obtained using Darcy’s law in the low $\nabla P$ regime, $k^\mathrm{F}$ and $k_1^\mathrm{F}$ are the permeability obtained using Forchheimer’s formula in the high $\nabla P$ regime.}
\label{tab:tab1}
\begin{tabular}{c c c c c c c}
\hline
Sample & Location (mm) & Porosity & Tortuosity & $k^\mathrm{D}$ ($10^{-13}\mathrm{m}^2$) & $k^\mathrm{F}$ ($10^{-13}\mathrm{m}^2$) & $k^\mathrm{F}_1$ ($10^{-9}\mathrm{m}^2$) \\
\hline
1 & (1.68, 20.35, 10.06) & 0.170 & 1.80 & 5.59 & 7.95 & 4.21\\
2 & (1.68, 16.76, 0) & 0.183 & 1.91 & 2.58 & 9.15 & 0.41\\
3 & (0, 14.45, 3.35) & 0.226 & 1.56 & 35.1 & 67.1 & 68.9\\
4 & (-3.35, 13.64, 6.73) & 0.162 & 1.91 & 7.99 & 14.9 & 5.77\\
5 & (0, 16.99, 0) & 0.180 & 1.65 & 14.3 & 24.9 & 17.5\\
6 & (-1.68, 18.67, 1.68) & 0.180 & 1.78 & 18.6 & 29.4 & 39.3\\

\hline
\end{tabular}
\end{table}

\begin{figure}[!htb]
     \centering
     \includegraphics[width=.7\linewidth]{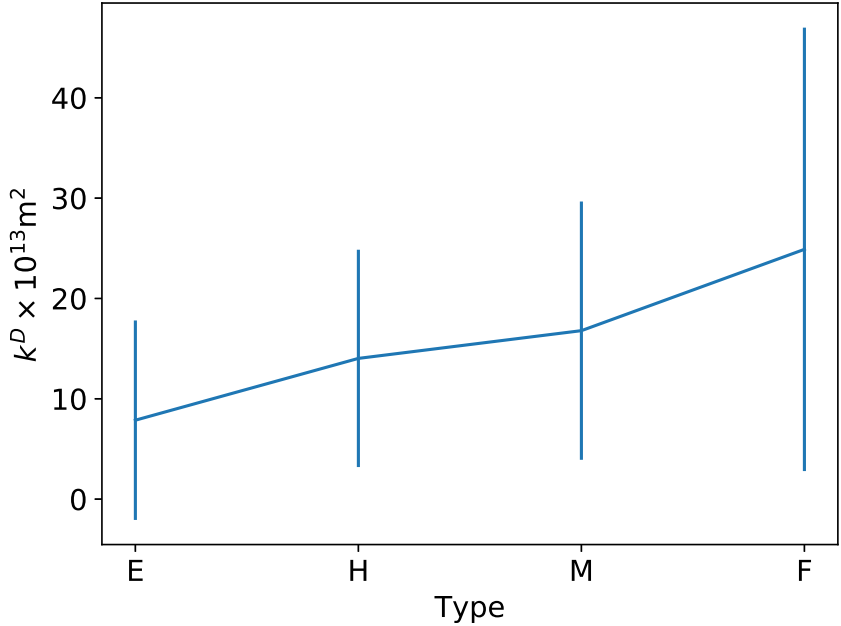}
     \caption{Permeability of different powder.}
     \label{fig:permeability_darcy_all}
\end{figure}

Percolation simulations are also run on samples from type E, M and F powder. The results are summarized in Tab.~\ref{tab:tab2}-\ref{tab:tab4}. The permeability $k^D$ for different types of powder at low pressure gradient is also shown in Fig.~\ref{fig:permeability_darcy_all}. It can be seen that, with the coffee particle becoming coarser the average permeability increases, and the variance of the permeability also increases.

\begin{table}
\centering
\caption{Summary of the properties of the coffee samples (type E powder). The parameter $k^\mathrm{D}$ is the permeability obtained using Darcy’s law in the low $\nabla P$ regime, $k^\mathrm{F}$ and $k_1^\mathrm{F}$ are the permeability obtained using Forchheimer’s formula in the high $\nabla P$ regime. Type E powder is the finest powder, therefore in some regions the flow is nearly clogged. Some parameters cannot be obtained and we mark them with dash in the table.}
\label{tab:tab2}
\begin{tabular}{c c c c c c c}
\hline
Sample & Location (mm) & Porosity & Tortuosity & $k^\mathrm{D}$ ($10^{-13}\mathrm{m}^2$) & $k^\mathrm{F}$ ($10^{-13}\mathrm{m}^2$) & $k^\mathrm{F}_1$ ($10^{-9}\mathrm{m}^2$) \\
\hline
1 & (0, 12.89, 0)                 & 0.141 & 2.61 & 1.34 & 2.29 & 0.45\\

2 & (3.35, 14.56, -5.02)          & 0.196 & 1.55 & 27.6 & 69.8 & 53.1\\

3 & (-10.04, 16.24, 3.35)         & 0.204 & 1.60 & 13.4 & 28.8 & 20.7\\

4 & (6.70, 17.91, -10.04)         & 0.132 & 2.71 & 0.27 & - & -\\

5 & (-1.67, 19.58, 3.35)          & 0.131 & - & 0 & - & -\\

6 & (-8.34, 20.42, 8.34)           & 0.164 & 1.76 & 4.59 & 7.40 & 2.17\\

\hline
\end{tabular}
\end{table}

\begin{table}
\centering
\caption{Summary of the properties of the coffee samples (type M powder). The parameter $k^\mathrm{D}$ is the permeability obtained using Darcy’s law in the low $\nabla P$ regime, $k^\mathrm{F}$ and $k_1^\mathrm{F}$ are the permeability obtained using Forchheimer’s formula in the high $\nabla P$ regime.}
\label{tab:tab3}
\begin{tabular}{c c c c c c c}
\hline
Sample & Location (mm) & Porosity & Tortuosity & $k^\mathrm{D}$ ($10^{-13}\mathrm{m}^2$) & $k^\mathrm{F}$ ($10^{-13}\mathrm{m}^2$) & $k^\mathrm{F}_1$ ($10^{-9}\mathrm{m}^2$) \\
\hline
1 & (5.02, 21.18, -6.70)                 & 0.193 & 1.66 & 38.5 & 93.5 & 79.4\\

2 & (0, 19.51, 0)                        & 0.149 & 2.36 & 1.74 & 3.17 & 0.46\\

3 & (10.06, 17.83, -10.06)               & 0.236 & 1.40 & 29.4 & 121.0 & 45.1\\

4 & (1.68, 16.99, 3.35)                  & 0.169 & 1.64 & 10.3 & 17.6 & 8.91\\

5 & (-5.03, 15.32, -3.35)                & 0.158 & 1.94 & 7.81 & 13.9 & 8.19\\

6 & (10.06, 13.64, 3.35)                 & 0.202 & 1.59 & 13.0 & 23.4 & 16.1\\

\hline
\end{tabular}
\end{table}

\begin{table}
\centering
\caption{Summary of the properties of the coffee samples (type F powder). The parameter $k^\mathrm{D}$ is the permeability obtained using Darcy’s law in the low $\nabla P$ regime, $k^\mathrm{F}$ and $k_1^\mathrm{F}$ are the permeability obtained using Forchheimer’s formula in the high $\nabla P$ regime.}
\label{tab:tab4}
\begin{tabular}{c c c c c c c}
\hline
Sample & Location (mm) & Porosity & Tortuosity & $k^\mathrm{D}$ ($10^{-13}\mathrm{m}^2$) & $k^\mathrm{F}$ ($10^{-13}\mathrm{m}^2$) & $k^\mathrm{F}_1$ ($10^{-9}\mathrm{m}^2$) \\
\hline
1 & (1.68, 20.35, 10.06)   & 0.193 & 1.41 & 66.6 & 213.6 & 161.4\\

2 & (0, 20.35, 0)          & 0.183 & 1.84 & 27.4 & 94.8 & 41.0\\

3 & (10.06, 18.67, -10.06) & 0.185 & 2.60 & 1.38 & 1.92 & 0.527\\

4 & (-10.06, 15.32, 10.06) & 0.211 & 1.62 & 20.7 & 66.2 & 24.1\\

5 & (0, 14.48, 3.35)       & 0.182 & 1.56 & 32.1 & 60.3 & 75.1\\

6 & (5.03, 12.80, -6.70)   & 0.171 & 2.41 & 1.19 & 1.98 & 0.43\\

\hline
\end{tabular}
\end{table}

From the results presented above we can see that the variance of permeability across the bed is quite large. Despite this, the trend of the average values of the permeability (see Fig.~\ref{fig:permeability_darcy_all}) is in agreement with our qualitative prediction based on the coarseness of the coffee particle: the more fines there are, and the larger the specific surface area, the smaller the permeability. Even though only a small number of samples is used for analysis, different powders are still able to show qualitative differences. Given enough computation resource quantitative comparison of extraction dynamics for different powders will be possible. 
 \subsection{Analysis of Post-extraction images}
 \label{sec:post-extraction}
\begin{figure}[hbt]
\centering
  \includegraphics[width=0.95\linewidth]{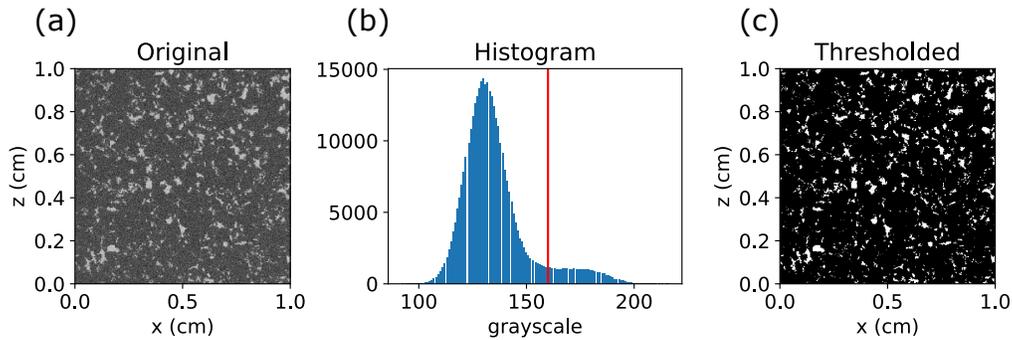}
  \caption{(a) Original microCT image of type H dark roast powder after extraction. (b) Histogram of grayscale of the pixels. The red line marks the manually-selected grayscale threshold. (c) Thresholded image.}
  \label{fig:thresholded_extracted}
  
\end{figure}

In order to study the link between changing microstructure and extraction, two additional illy iperespresso capsules are further extracted, and the post-extraction microCT imaging are analysed. These capsules are extracted using an X7.1 Iperespresso - Capsules Coffee Machine at water temperature $97 \pm 3^\circ$C and pressure $13\pm1$ bar. The capsules are scanned with microCT before and after the extraction so that a comparison can be made to investigate the morphological change caused by the extraction.
Because we do not know the porosity of the extracted coffee cakes \textit{a priori}, the segmentation threshold cannot be determined by matching the porosity. However, since after wetting the distinction between coffee grains and intergranular pores are much more significant, we can manually select the grayscale threshold to capture the phase boundary. An example of microCT imaging of the extracted dark roast capsule is shown in Fig.~\ref{fig:thresholded_extracted} (a). As can be seen, the boundary between the grains and the void is much clearer here than that in Fig.~\ref{fig:thresholded} (a). Fig.~\ref{fig:thresholded_extracted} (b) also shows that the distribution of the grayscale in this microCT image is bimodal. Therefore we can select the threshold manually to binarise the microCT image: we choose the threshold to be the horizontal center of the two distribution peaks. A threshold selected in this way makes the segmentation least sensitive to the threshold value. The segmentation result of Fig.~\ref{fig:thresholded_extracted} (a) is shown in Fig.~\ref{fig:thresholded_extracted} (c).

\begin{figure}[hbt]
\centering
  \includegraphics[width=0.5\linewidth]{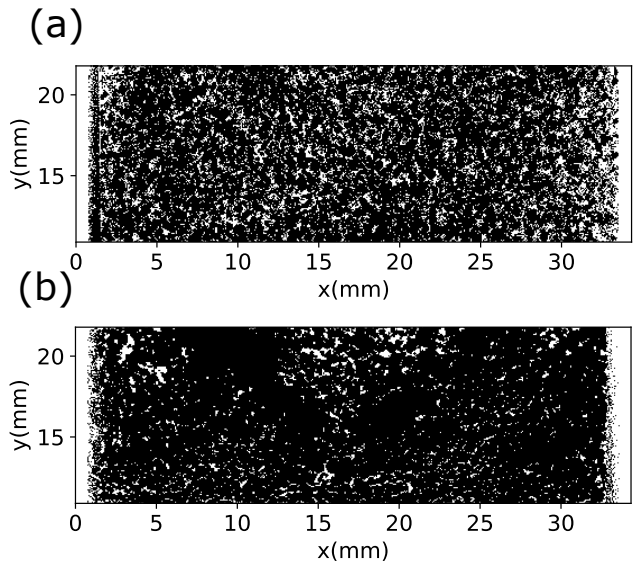}
  \caption{(a) Thresholded microCT image of type H medium roast powder, (a)  before extraction, (b) 3 days after extraction.}
  \label{fig:classico_imaging}
\end{figure}

\begin{figure}[hbt]
\centering
  \includegraphics[width=0.9\linewidth]{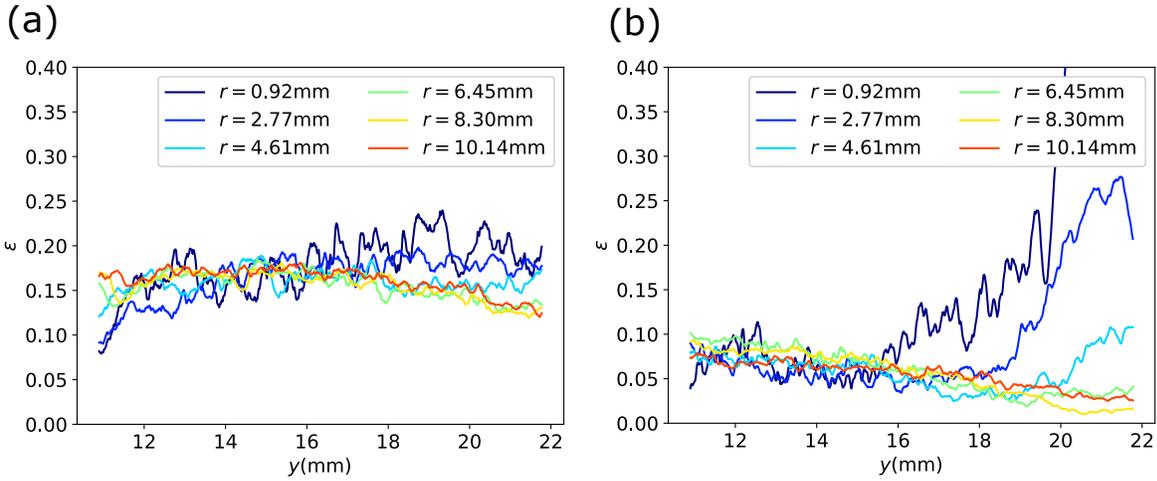}
  \caption{Porosity of the type H medium roast powder, (a) before extraction, (b) 3 days after extraction.}
  \label{fig:classico_porosity}
\end{figure}

\begin{figure}[hbt]
\centering
  \includegraphics[width=0.5\linewidth]{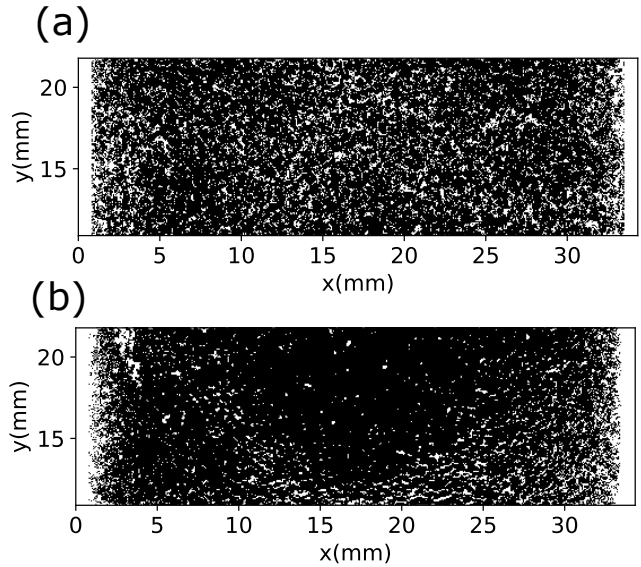}
  \caption{(a) Thresholded microCT image of type H dark roast powder, (a)  before extraction, (b) right after extraction.}
  \label{fig:intenso_imaging}
\end{figure}

\begin{figure}[hbt]
\centering
  \includegraphics[width=0.9\linewidth]{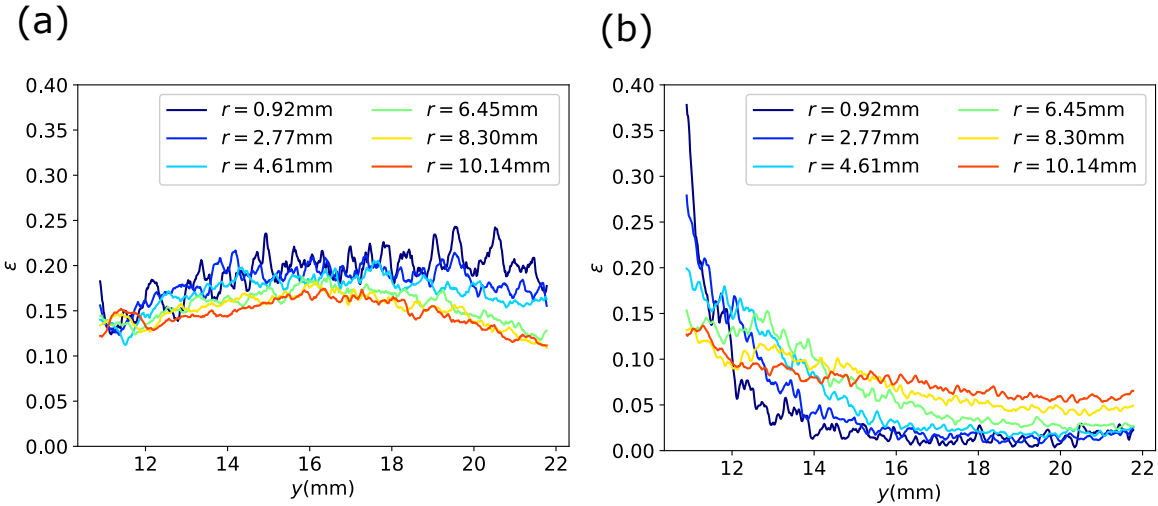}
  \caption{Porosity of the type H dark roast  powder, (a)  before extraction, (b) right after extraction.}
  \label{fig:intenso_porosity}
\end{figure}

A capsule with type H medium roast powder and a capsule with type H dark roast powder are scanned before and after extraction. The comparison of the binarized microCT images before and after extraction are shown in Fig.~\ref{fig:classico_imaging} and \ref{fig:intenso_imaging}. The comparison of the porosity before and after extraction are presented in Fig.~\ref{fig:classico_porosity} and \ref{fig:intenso_porosity}. It can be seen that for all the capsules, after extraction the coffee cakes become less porous, the porosity decreases significantly. That is because after wetting the coffee grains swell\cite{Mateus2007,Hargarten2020} and stick to their neighbours, liquid fills the intragranular pores and also lots of intergranular pores. The microCT cannot properly distinguish the liquid phase from the solid phase, resulting in an underestimation of the porosity. However, we focus on the relative spatial porosity variation in the coffee cakes rather than the absolute porosity values. In Fig.~\ref{fig:classico_porosity} the porosity of the type H medium roast powder before extraction and 3 days after extraction are shown. Before extraction the porosity is relatively homogenous. On the opposite, 3 days after the extraction, the porosity at the top center becomes much larger than in other regions of the matrix. This is probably due to the drying of the matrix,  i.e. the drying is faster at the top center. However, for the regions away from the center the porosity is always larger on the bottom (outlet) than at the top (inlet). This phenomenon is counterintuitive since we would expect the gravity will drive the residual water and other fragments to the bottom making the lower part of the matrix less porous. During extraction the applied external pressure should also have consolidated the matrix more efficiently on the bottom. But surprisingly, after extraction we observe larger porosity in the lower part (outlet) than in the upper part (inlet). To exclude the effect of drying, in Fig.~\ref{fig:intenso_porosity} the porosity of the type H dark roast powder before extraction and right after extraction are shown. Without the further complexity induced by the drying process, we see clearly that the porosity is always higher on the bottom after the extraction.
More experiments have been performed with the post-extraction matrix being scanned 1 hr or 24 hrs after the extraction, we confirm that the porosity profiles always have an apparent decreasing trend from the bottom to the top of the matrix.

Besides the drying process and consolidation, other physical processes that could also affect the geometry of the matrix are fragment migration\cite{Petracco1993,Ellero2019} and grain swelling\cite{Hargarten2020,Maille2021}. In our prior works, we have modelled both the fine migration\cite{Ellero2019} and particle swelling\cite{Mo2022} using SPH. But the fine migration will tend to lead to lower porosity near the bottom as the filter on the bottom prevents the fines from leaving the capsule. On the other hand, if swelling process is initiated earlier at the top it will cause the matrix to be less porous at the top. This is indeed the case, since the water is infused from the top to the bottom. However, the difference of porosity caused by this effect is expected to be minor after extraction, as the infusion process is fast ($\approx 1\mathrm{s}$), and the swelling process of all grains will eventually saturate in about 4min\cite{Hargarten2020}, while we still see the porosity difference 1hr after the extraction.
 
The most probable cause for the observed spatial porosity distribution is the heterogenous erosion process. Scanning electron microscopy observation has shown that immersed coffee grain has gas entrapped\cite{Mateus2007}. Coffee particles are not fully wetted during extraction due to the abundant gas entrapped in the intra-grain pores. Therefore, some part of the coffee particles remain dry and brittle in the extraction process. And for brittle material, the failure criterion is usually pressure-dependent. Note that the pressure is also quite non-uniform in the coffee matrix as a result of the filtration process. The heterogeneity of the post-extraction porosity is probably correlated to the pressure drop across the matrix through the pressure-dependent failure criterion of eroding material. As a matter of fact, in our previous numerical simulation study\cite{Mo2021}, we propose a bottom-up mesoscopic erosion model in the framework of SPH that incorporates the Mohr-Coulomb yield criterion\cite{Christensen2013,Labuz2012mohr} and the simple shear-erosion model\cite{Chiu2020JFM,Matias2020}. Simulations using this model show that heterogeneity in filtration direction does occur as a result of the pressure-dependent erosion: an initially relatively uniform porosity distribution becomes highly non-uniform after filtration with higher porosity on the bottom (the flow outlet) and lower porosity at the top (the flow inlet). The non-uniformity is mainly controlled by the internal friction angle of the eroding material. The post-extraction porosity profiles reported here are in qualitative agreement with the prediction of our erosion model\cite{Mo2021}. This suggests that our proposed model\cite{Mo2021} is suitable to capture the special characteristics of coffee cake erosion.

\section{Summary and conclusions}
\label{sec:conclusions}
To summarize, we use high-resolution X-ray microCT technique to capture the microscopic details of coffee matrices at particle-level. The 3D reconstructured data are then used to build an SPH-based digital twin for the percolation process. 

We explore the link between microstructure and flow properties from two perspectives: 1) The permeability and tortuosity. Using 3D microCT reconstruction data combined with SPH percolation simulations, both the permeability and tortuorsity of the matrices can be determined. We found that the inertial effects are actually quite significant at normal pressure gradient of espresso brewing. There have been some reports in literatures showing that larger pressure gradient seems to lead to lower coffee matrix permeability. We demonstrate here that this can be explained by considering the inertial effect. We also examine several types of coffee powder, the permeability and tortuosity are measured systematically using SPH simulation. Our results show that the variance of the permeability at different location of the same matrix is quite large, and the coarser the powder the larger the variance. Despite this, the average of the permeability steadily increases as the powder becomes coarser. 2) The post-extraction microstructure. We analyzed the microCT images of post-extraction coffee matrices, and found that a decreasing porosity profile (from the bottom outlet to the top inlet) always develops after extraction. This counterintuitive phenomenon can be explained using a pressure-dependent erosion model proposed in our prior paper\cite{Mo2021}. It is suggested that the erosion of coffee powder is to some extent pressure-dependent. This pressure-dependence probably originates from the gas entrapped inside the coffee particles: grains not fully wetted will remain sufficiently brittle, and the failure of brittle material is usually pressure dependent.
Another physical scenario consistent with this pressure-dependent erosion model is connected to the presence of individual fines adhering electrostatically to the coarse particles, and being removed by the flow when their superficial attraction force is overcome by hydrodynamic normal-tangential forces.

To summarize, using the microCT and SPH simulations we have obtained two key findings including the identification of the Darcy-Forchheimer transition and the heterogeneous porosity profile in the filtration direction after extraction. The Darcy-Forchheimer transition suggests that the decrease of permeability under higher extraction pressure found in many experiments can be attributed to inertial effects. It helps to clarify the discrepancy in experiments from different research groups and provides guidance for new measurement experiments. The post-extraction heterogeneity in the filtration direction indicates that the erosion of coffee particles can be pressure-dependent. It highlights the role played by the erosion mechanism of coffee grains under hydrodynamic forces.

We mainly explored the link between the microstructure obtained from the microCT and the flow properties in a qualitative way. Constrained by the computational resource, a 3D model of a full realistic size can not be simulated to establish quantitative links. Nevertheless, our results show that microCT scan can provide us many microscopic details of a coffee matrix. These information is critical if we want to accurately model the percolation process. Moreover, by comparing the pre- and post-extraction scans we can obtain some crucial information about how the geometry changes during extraction process. This can in turn reveal some details of how the interconnected complex physical processes develop during coffee extraction, and helps us to propose better modelling methods to account for these processes.

\section*{Acknowledgement}

This research is supported by the illycaffè S.p.A.\ through the project “Modelling and simulating espresso coffee extraction at the mesoscopic scales”. Financial support from the BERC 2022-2025 program and by the Spanish State Research Agency through BCAM Severo Ochoa Excellence Accreditation CEX2021-001142-S/MICIN/AEI/10.13039/501100011033 and through the project PID2020-117080RB-C55 (“Microscopic foundations of soft matter experiments: computational nano-hydrodynamics” - acronym “Compu-Nano-Hydro”) are also acknowledged. 

The microCT work was supported by the Advanced Imaging Materials (AIM) core facility (EPSCR Grant No. EP/M028267/1), and the European Social Fund (ESF) through the European Union Convergence programme administered by the Welsh Government (80708).
\section*{Competing interests statement}
The co-authors affiliated to illycaffè S.p.A.\ declare that they have no known competing financial interests or personal relationships that could have appeared to influence the work reported in this paper. All other authors also declare no competing interests. 
\section*{Data availability}
The datasets used and/or analysed during the current study available from the corresponding author on reasonable request.
\clearpage
\bibliographystyle{unsrt}
\bibliography{main}
\end{document}


\title{Supplementary Material: Exploring the link between coffee matrix microstructure and flow properties using combined X-ray microtomography and smoothed particle hydrodynamics simulations}
\author[1,2]{Chaojie Mo\footnote{cmo@bcamath.org}}
\author[3]{Richard Johnston}
\author[4]{Luciano Navarini}
\author[4]{Furio Suggi Liverani}
\author[1,5,6]{Marco Ellero}
\affil[1]{Basque Center for Applied Mathematics (BCAM), Alameda de Mazarredo 14, 48009 Bilbao, Spain}
\affil[2]{Aircraft and Propulsion Laboratory, Ningbo Institute of Technology, Beihang University, Ningbo 315100, P. R. China}
\affil[3]{Faculty of Science and Engineering, Swansea University, Swansea, SA1 8EN, UK}
\affil[4]{Illycaffè S.p.A, Via Flavia 110, Trieste 34147, Italy}
\affil[5]{Zienkiewicz Centre for Computational Engineering (ZCCE), Swansea University, Bay Campus, Swansea SA1 8EN, UK}
\affil[6]{IKERBASQUE, Basque Foundation for Science, Calle de María Díaz de Haro 3, 48013 Bilbao, Spain}
\date{}
\maketitle
\section{Smoothed Particle Hydrodynamics formalism}
In our model the flow is governed by the Navier-Stokes equations. 
To solve them, we employ the smoothed particle hydrodynamics (SPH) method\cite{Monaghan_SPH_1992, Ellero2007}, which is a particle-based hydrodynamics approach. SPH is derived through a Lagrangian discretisation of the Navier-Stokes equations.
We employ an SPH version which conserves angular momentum \cite{Mueller_SDPD_2015}, as it can be crucial for some problems
	\cite{Hu_AMC_2006,Gotze_RAM_2007,Vazquez2016jnnfm}. In SPH, each particle can be considered as a small fluid volume (or Lagrangian discretisation point) characterised by a position $\bm{r}_i$, velocity $\bm{v}_i$, and mass $m_i$. In addition, each SPH particle possesses a spin angular velocity $\bm{\psi}_i$ and moment of inertia $I_i$ introduced for the enforcement of angular momentum conservation \cite{Mueller_SDPD_2015}. 
	
SPH particles $i$ and $j$ interact through three pairwise forces, including conservative $\bm{F}_{ij}^C$, dissipative $\bm{F}_{ij}^D$, and rotational $\bm{F}_{ij}^R$ forces given by 
	\begin{equation}
	 \begin{gathered}
	   \bm{F}_{ij}^C = \left(\frac{P_i}{d_i^2} + \frac{P_j}{d_j^2} \right)F_{ij} \bm{r}_{ij}, \\
	   \bm{F}_{ij}^D = - \gamma_{ij}\left[\bm{v}_{ij} + (\bm{e}_{ij}\cdot\bm{v}_{ij})\bm{e}_{ij}\right], \\
	   \bm{F}_{ij}^R = - \gamma_{ij}\frac{\bm{r}_{ij}}{2} \times (\bm{\psi}_i + \bm{\psi}_j),
	 \end{gathered}
	 \label{eq:forces}
	\end{equation}
	where $\bm{r}_{ij} = \bm{r}_i - \bm{r}_j$, $\bm{v}_{ij} = \bm{v}_i - \bm{v}_j$, and $\bm{e}_{ij} = \bm{r}_{ij}/r_{ij}$. Particle number density $d_i$ is computed as $d_i=\sum_j W_{ij}$ using a smoothing kernel function $W_{ij} = W(r_{ij})$ that vanishes beyond a cutoff radius $r_c$ and defines a non-negative function $F_{ij}$ through the equation $\nabla_i W_{ij} = -\bm{r}_{ij} F_{ij}$. Then, particle mass density is given by $\rho_i = m_i d_i$. The pressure $P_i$ is determined by the equation of state (EoS) $P_i = P_0(d_i/d_0)^\nu-P_\mathrm{b}$, where $d_0$ is the average number density, $P_0$ and $\nu$ are parameters controlling the sound speed $c=\sqrt{P_0 \nu/d_0}$, and $P_\mathrm{b} $ relates to the background pressure. Furthermore, $\gamma_{ij}$ is a force amplitudes defined as
	\begin{equation}
	  \gamma_{ij} = \frac{20\eta}{7} \frac{F_{ij}}{d_i d_j},
	\end{equation}
	where $\eta$ is the fluid dynamic viscosity.    
	
The evolution of particle positions, translational and angular velocities is obtained by integration of the following equations 
	of motion 
	\begin{equation}
	  \begin{gathered}
	    \dot{\bm{r}}_i = \bm{v}_i,  \\
	    m_i\dot{\bm{v}}_i = \sum_j \bm{F}_{ij} = \sum_j (\bm{F}_{ij}^C + \bm{F}_{ij}^D + \bm{F}_{ij}^R), \\ 
	    \dot{\bm{\psi}}_i = \frac{1}{2I_i}\sum_j \bm{r}_{ij}\times \bm{F}_{ij}, 
	  \end{gathered}
	  \label{eq:sdpd}
	\end{equation}
	using the velocity-Verlet algorithm \cite{Allen_CSL_1991}. 
	
	In this work, the smoothing kernel is represented by the quintic spline kernel function \cite{Ellero2005}
	\begin{equation}
    W(q) = w_0	
	\begin{cases}
      (3-q)^5-6(2-q)^5+15(1-q)^5, & 0 \le q<1 \\
      (3-q)^5-6(2-q)^5, & 1 \le q<2 \\
      (3-q)^5, & 2\le q < 3 \\
      0, & q\ge 3
	\end{cases}
	\end{equation}
    where $q=r/h$ and $w_0=1/(120\pi h^3)$ in three dimensions (3-D), $w_0=7/(478\pi h^2)$ in two dimensions (2-D), and $h$ is the smoothing length $h=r_c/3$.
    
    The basic parameters used in our simulations are summarized in table \ref{tab:sph_parameters}.
    \begin{table*}
	\centering
	\caption{Basic parameters for the SPH simulations.}
	\label{tab:sph_parameters}
	\begin{tabular}{ l l l l}
	\hline	
	\textbf{Parameters} & \textbf{Values} & \textbf{Dimensions} & \textbf{Values in SI units}\\
	\hline
	Cutoff radius $r_c$ & $1.0$ & $\mathrm{L}$ & $51.7 \mathrm{\mu m}$\\
	Mass density $\rho$  & $30.0$ & $\mathrm{ML^{-3}}$ & $1000 \mathrm{kg/m}$\\
	Dynamic viscosity $\eta$ & $2$ & $\mathrm{ML^{-1}T^{-1}}$ &  $1 \mathrm{mPa\cdot s}$\\
	Average number density $d_0 $ & $30$ & $\mathrm{L^{-3}}$ & $217490\mathrm{mm}^{-3}$\\
	SPH particle mass $m$ & 1 & $\mathrm{M}$ & $4.60\times 10^{-12} \mathrm{kg}$\\
	Moment of inertial of SPH particles $I$ & 1 & $\mathrm{ML^2}$ & $1.23\times 10^{-20}\mathrm{kg\cdot m^2}$ \\
	$P_0$ in the EoS & 6400 & $\mathrm{ML^{-1}T^{-2}}$ & $18.0\mathrm{kPa}$\\
	Hydrostatic pressure $P_0 - P_\mathrm{b}$ & 80 & $\mathrm{ML^{-1}T^{-2}}$ & $225\mathrm{Pa}$\\
	Exponent in the EoS $\nu$ & $7$ & 1 & 7 \\
	Time step $\Delta t$ & 0.001 &$\mathrm{T}$ & $0.18\mathrm{ms}$\\
	\hline
	\end{tabular}
	\end{table*}
\section{Enlarged vertical sections of the microCT imaging for different powders}
\begin{figure}[hbt]
\centering
  \includegraphics[width=1.0\linewidth]{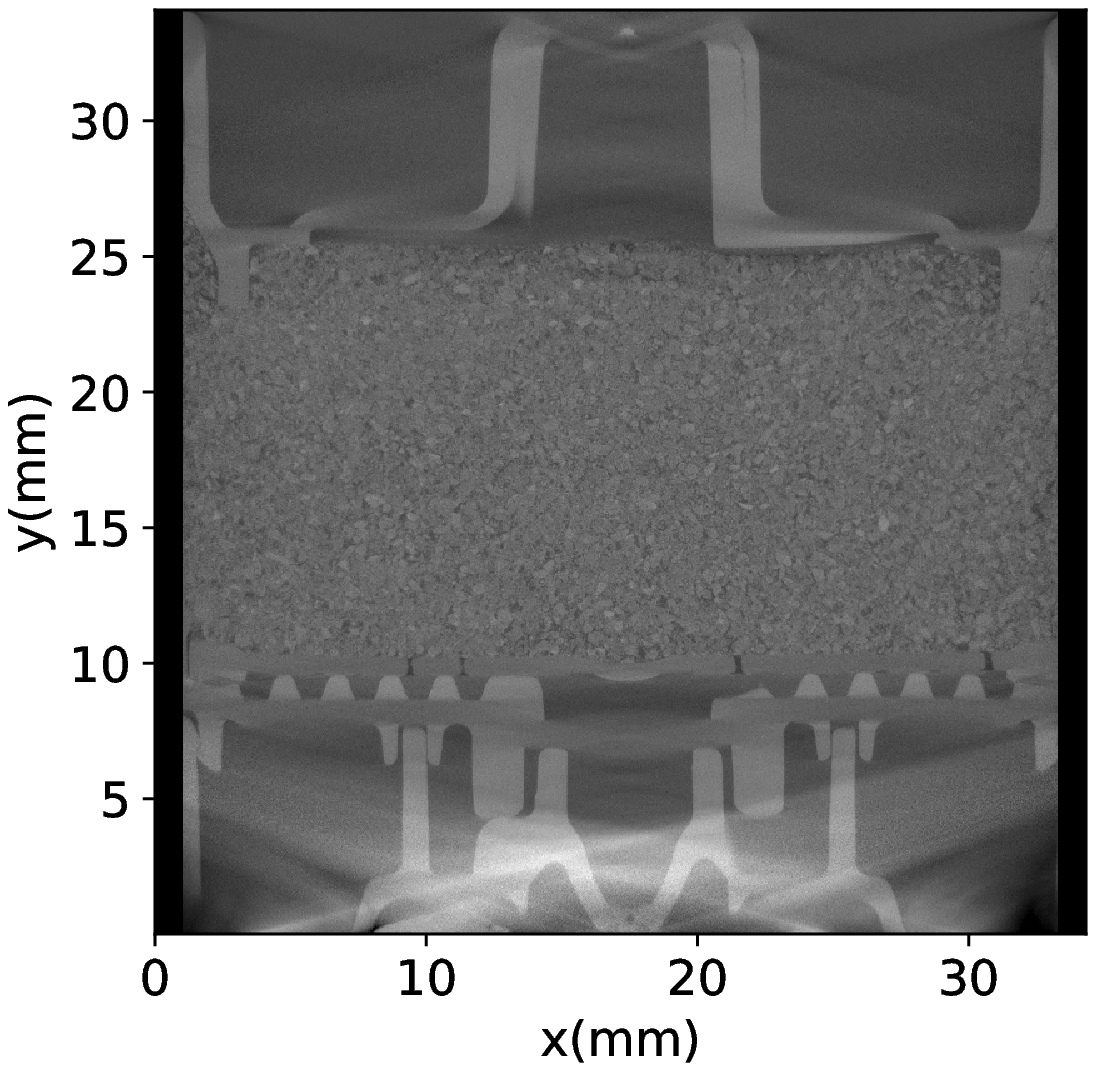}
  \caption{An enlarged vertically-cut section of the microCT imaging for the type E coffee powder.}
  \label{fig:typeE_CT}
\end{figure}

\begin{figure}[hbt]
\centering
  \includegraphics[width=1.0\linewidth]{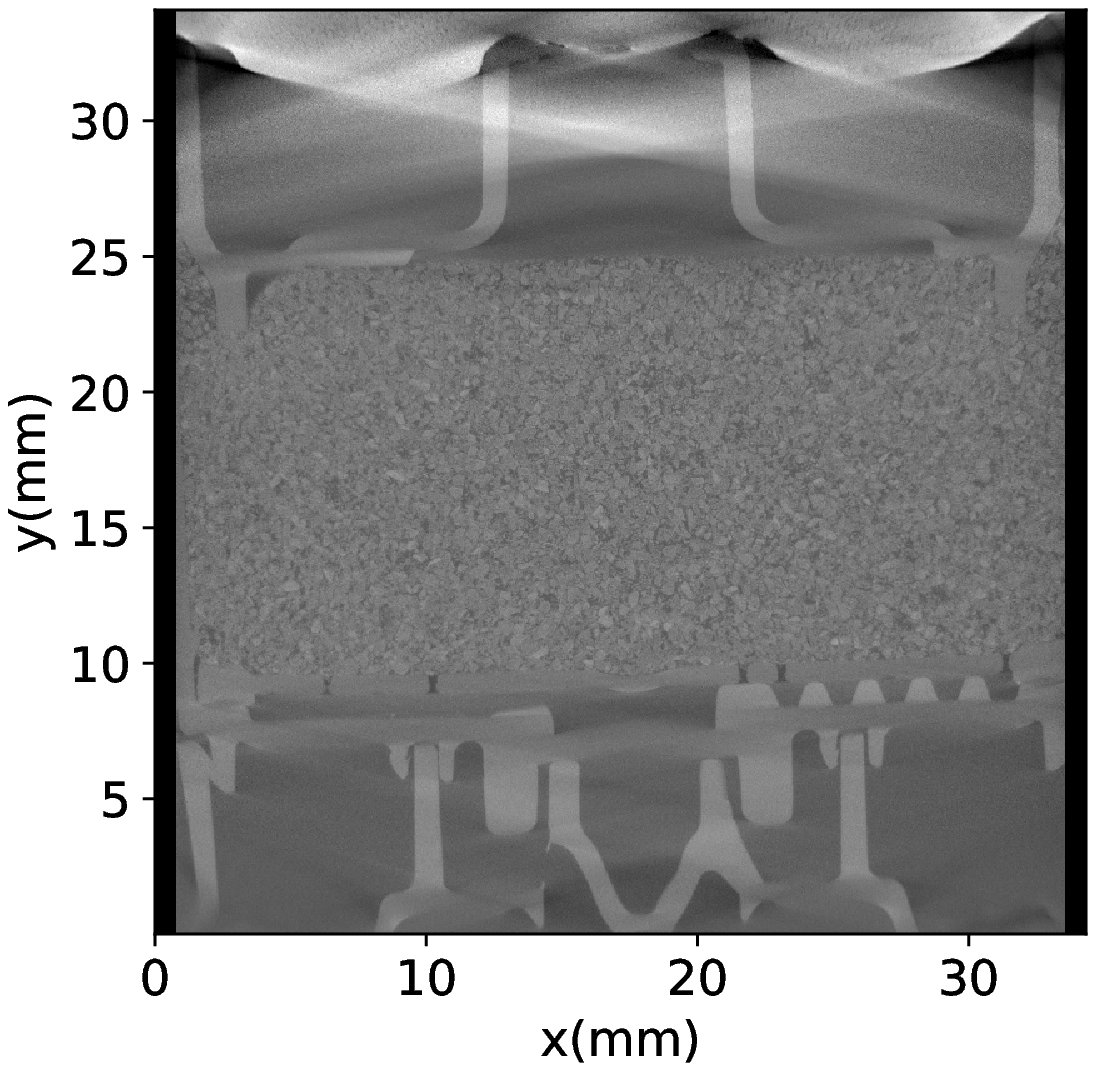}
  \caption{An enlarged vertically-cut section of the microCT imaging for the type H coffee powder.}
  \label{fig:typeH_CT}
\end{figure}

\begin{figure}[hbt]
\centering
  \includegraphics[width=1.0\linewidth]{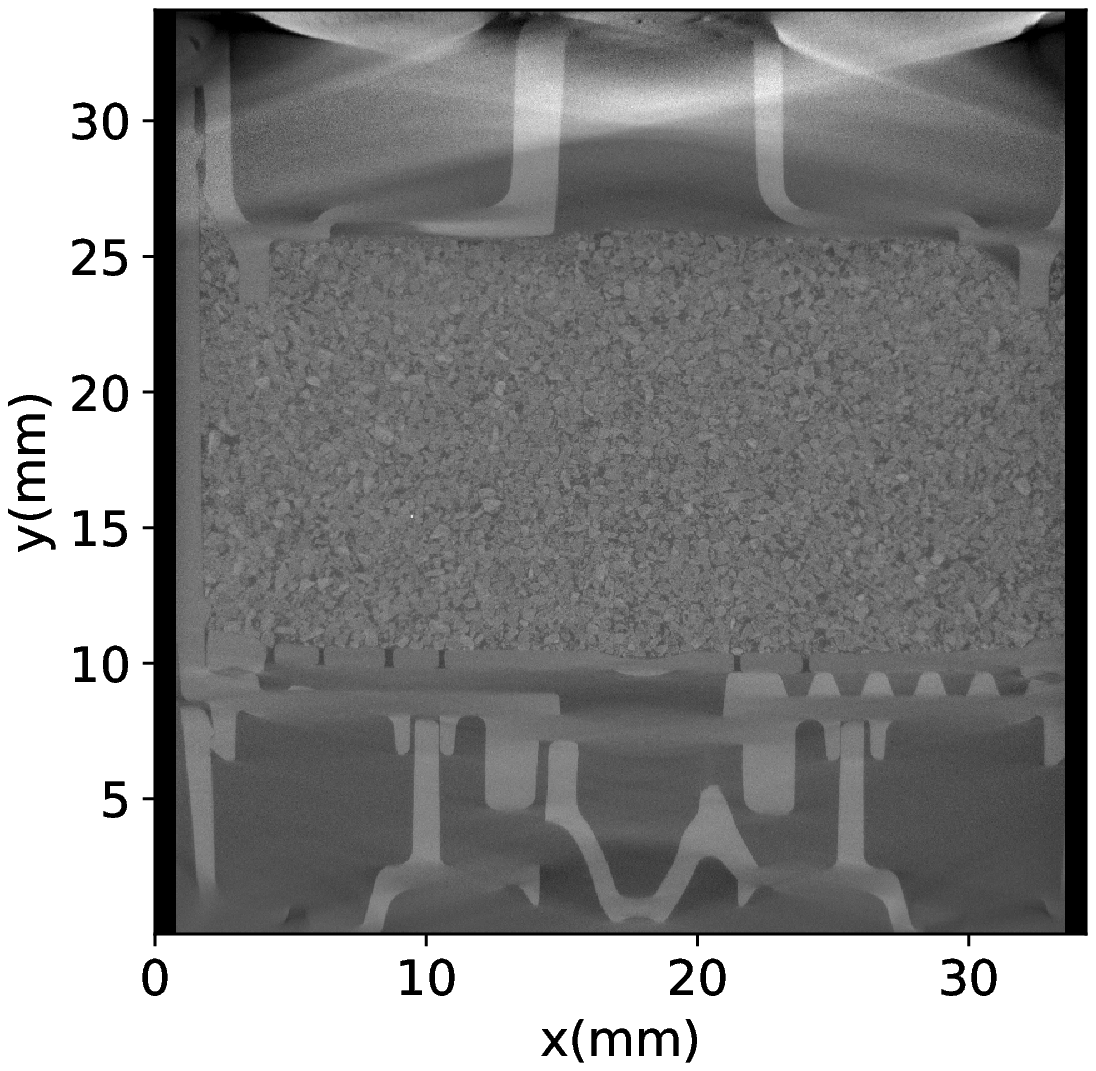}
  \caption{An enlarged vertically-cut section of the microCT imaging for the type M coffee powder.}
  \label{fig:typeM_CT}
\end{figure}

\begin{figure}[hbt]
\centering
  \includegraphics[width=1.0\linewidth]{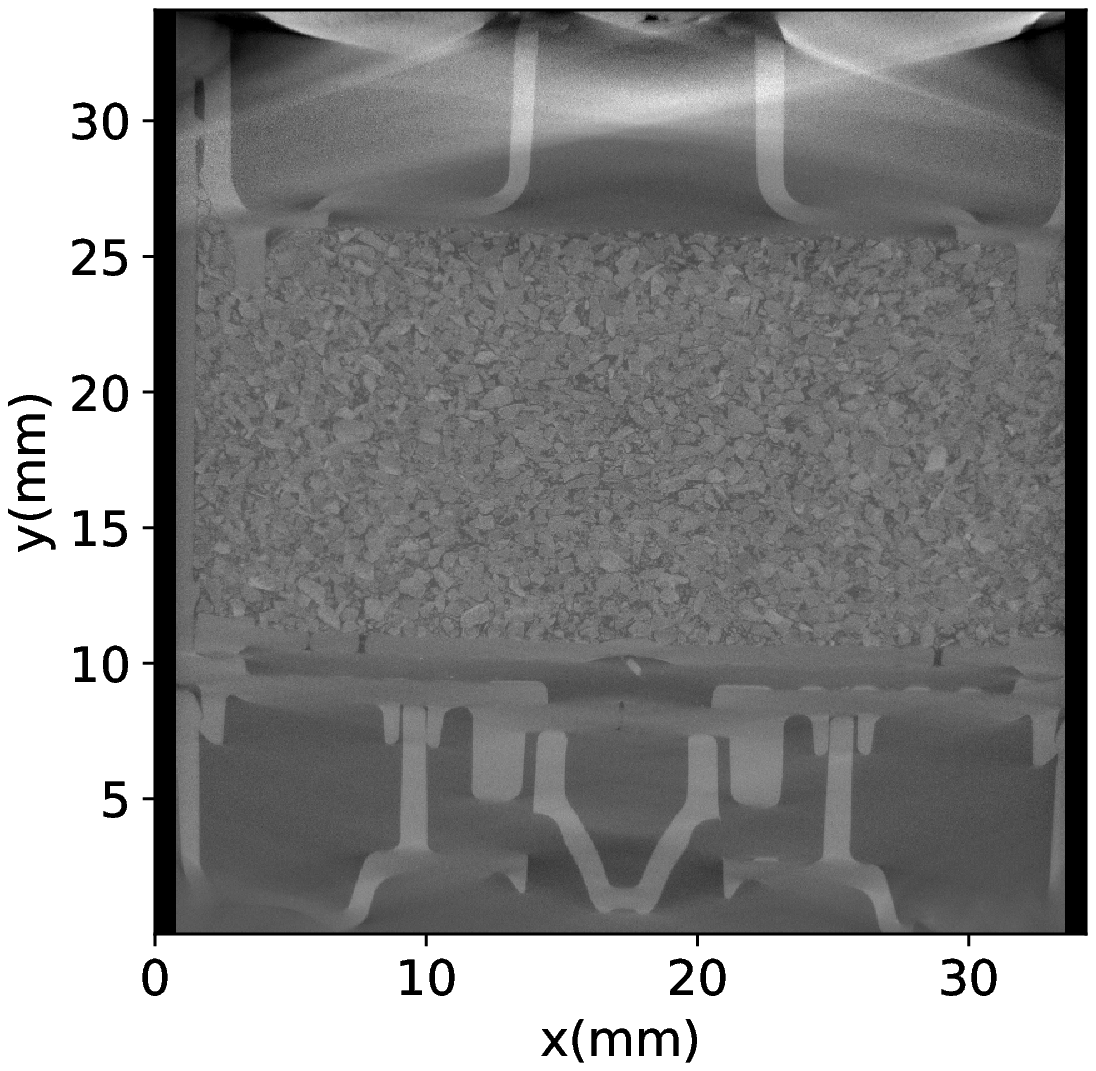}
  \caption{An enlarged vertically-cut section of the microCT imaging for the type F coffee powder.}
  \label{fig:typeF_CT}
\end{figure}

We show enlarged vertical digital sections of the microCT/XRM imaging data for different powders in Fig.~\ref{fig:typeE_CT} (type E), Fig.~\ref{fig:typeH_CT} (type H), Fig.~\ref{fig:typeM_CT} (type M), and Fig.~\ref{fig:typeF_CT} (type F). Both the coffee grains and the plastic/metal structure skeletons of the capsules are clearly visible in these images. It is also discernible that the coffee particles get coarser from type E to type F.
\bibliographystyle{unsrt}
\clearpage
\bibliography{main}